\documentclass[structabstract]{aa}  
\usepackage{graphicx}
\usepackage{txfonts}
\usepackage{natbib}
\bibpunct{(}{)}{;}{a}{}{,}

\begin{document}

\title{Photospheric activity, rotation and magnetic interaction in LHS~6343~A}

\author{E.~Herrero\inst{\ref{inst1}} \and A.~F.~Lanza\inst{\ref{inst2}} \and I.~Ribas\inst{\ref{inst1}} \and 
C.~Jordi\inst{\ref{inst3}} \and J.~C.~Morales\inst{\ref{inst1},\ref{inst3}}
} 
\institute{Institut de Ci\`{e}ncies de l'Espai (CSIC-IEEC), Campus UAB, 
Facultat de Ci\`{e}ncies, Torre C5 parell, 2a pl, 08193 Bellaterra, 
Spain, \email{eherrero@ice.cat, iribas@ice.cat, morales@ice.cat}\label{inst1}
\and
INAF - Osservatorio Astrofisico di Catania, via S. Sofia, 78, 95123 Catania, Italy,
\email{nuccio.lanza@oact.inaf.it}\label{inst2}
\and
Dept. d'Astronomia i Meteorologia, Institut de Ci\`{e}ncies del Cosmos (ICC),
%\footnote{Associated with the Instituto de Ciencias del Espacio. Consejo Superior de Investigaciones Cientificas.}
Universitat de Barcelona (IEEC-UB), Mart\'{i} Franqu\`{e}s 1, E08028 Barcelona, Spain, 
\email{carme.jordi@ub.edu}\label{inst3}
}
\date{Received <date> /
Accepted <date>}

\abstract
% context heading
{The Kepler mission has recently discovered a brown dwarf companion transiting 
one member of the M4V+M5V visual binary system LHS 6343 AB with an orbital period of 12.71 days.}
% aims heading
{The particular 
interest of this transiting system lies in the synchronicity between the transits of 
the brown dwarf C component and the main modulation observed in the light curve, which is assumed 
to be caused by rotating starspots on the A component. We model the  activity 
of this star by deriving maps of the active regions that allow us to study stellar rotation
 and the possible interaction with the brown dwarf companion.}
% methods heading
{{ An average transit profile was derived, and the photometric perturbations due to spots occulted during  transits are removed to derive more precise transit parameters. } We applied a maximum entropy spot model to fit the out-of-transit optical modulation as observed by 
Kepler during an uninterrupted interval of $\sim500$ days. It assumes that stellar active 
regions consist of cool spots and bright faculae whose visibility is modulated by 
stellar rotation.}
% results heading
{ { Thanks to the extended photometric time series,  we refine the determination of the transit parameters and find evidence of spots that are occulted by the brown dwarf during its transits. }
The modelling of the out-of-transit light curve of LHS 6343 A reveals 
several starspots rotating with a slightly longer period than  
the orbital period of the brown dwarf, { i.e., $13.13 \pm 0.02$~days}. No signature attributable to  
differential rotation is observed. We find  evidence of a persistent active longitude on the M dwarf preceding the sub-companion point by $\sim100^{\circ}$ and lasting for 
at least $\sim500$~days. This can be relevant for understanding how 
magnetic interaction works in low-mass binary  and star-planet systems. }{}

\keywords{stars: magnetic fields -- stars: late-type -- stars: activity --
  stars: rotation -- binaries: eclipsing -- brown dwarfs} 
\maketitle

\section{Introduction}

Kepler is a photometric space mission devoted to searching for planets 
by the method of transits and asteroseismology. The recently released public data contain a large amount of 
information useful for characterizing stellar pulsations and modulations caused by starspots 
in solar-like and low-mass stars.

M-type stars are particularly interesting targets in exoplanet searches since Earth-like 
planets in their habitable zones  are expected to lie within the detection capabilities 
of some of the currently working instruments. However, the effects of magnetic activity in these 
stars may sometimes mask or even mimic the signal of a transiting exoplanet \citep[e.g., ][]{BonomoLanza08,Barnes11}. High-precision photometry from space 
offers the opportunity to study the signature of spots 
and faculae during large uninterrupted intervals of time, so it represents the best 
way to improve our knowledge of activity patterns in low-mass stars in order to further optimize exoplanet searches.

Among the M-type stars observed by Kepler, a particularly interesting object is LHS 6343 AB, a visual binary system consisting of an M4V and an M5V star at a projected distance of $\sim40.2$ AU. \cite{2011ApJ...730...79J} report the discovery of a brown dwarf companion (hereafter LHS~6343~C) with a mass of $62.7\pm2.4 M_{\rm Jup}$ transiting component A of the system every 12.7138 days. The primary star, LHS~6343~A, shows almost no evidence of chromospheric activity and presents a modulation of very low amplitude in the Kepler optical light curve (see Sects.~\ref{sec_obs} and~\ref{sub_transit}). The modulation remains stable for more than 500 days and has a period of 12.71$\pm$0.28 days, which is synchronized with the orbital motion of the companion. The large separation between the A and C components makes  tidal interaction particularly weak in this system. Some other known synchronous systems are \object{CoRoT-3} \citep{Deleuiletal08,Triaudetal09}, \object{CoRoT-4} \citep{Lanzaetal09b}, and $\tau$ Bootis \citep{Catalaetal07,Walkeretal08,Donatietal08}, although in the last two cases the star is orbited by a massive planet, not by a brown dwarf (hereafter BD). Since they  are  F-type dwarf stars, it has been suggested that the close-in brown dwarf or the planets may have synchronized their thin outer convective envelopes during the main-sequence lifetimes of the systems \citep{Donatietal08}. 

In this work we study the  photospheric activity of LHS~6343~A by applying the spot modelling approach already used for other objects, such as \object{CoRoT-4} \citep{Lanzaetal09b} or \object{CoRoT-6} \citep{Lanzaetal11}. This allows us to map the longitudinal distribution and evolution of spots in a reference frame co-rotating with the companion orbital period. As we discuss in Sect.~\ref{sub_activ}, we find an overall migration of the spot pattern, indicating a rotation period of 13.137$\pm$0.011 days for LHS~6343~A, and a persistently active longitude, locked at a phase difference of $\sim100^{\circ}$ from the sub-companion point. Starspots are systematically enhanced when they approach and cross that longitude during their migration. This leads us to discuss possible magnetic interactions between LHS~6343~A and its brown dwarf companion (cf. Sect.~\ref{sec_discus}).

\section{Observations}
\label{sec_obs}

About 156,000 targets are being continuously monitored by the Kepler space telescope in a $10^{\circ} \times 10^{\circ}$ field near the constellations of Lyra and Cygnus \citep{Kochetal10,Boruckietal10}. Photometry is continuously acquired for time intervals up to $\sim 90$~days, called "quarters" in Kepler jargon, after which a re-orientation of the spacecraft is required to keep the solar arrays pointed toward the Sun. This maneuver implies a rotation of the field of view on the focal plane making the image of a given star to fall on a different CCD. As a consequence, a jump appears in the raw time series at the end of each quarter.  
 Kepler acquires photometry with a so-called long-term cadence of $\sim 30$ minutes,  except for a subset of selected targets for which the exposure is reduced to about one minute. 
Long-cadence data have been made publicly available as part of the first quarters data release (Q0-Q6) and their reduction is described in \citet{Jenkinsetal10a,Jenkinsetal10b}.

%%%%%%%%%%%%%%%%%%%%%%%%%%%%%%%%%%%%%%%%%%%%%%%%%%%%%%%%%%%%%%%%%
\begin{figure*}[t]
\centerline{
\includegraphics[width=14cm,angle=-90]{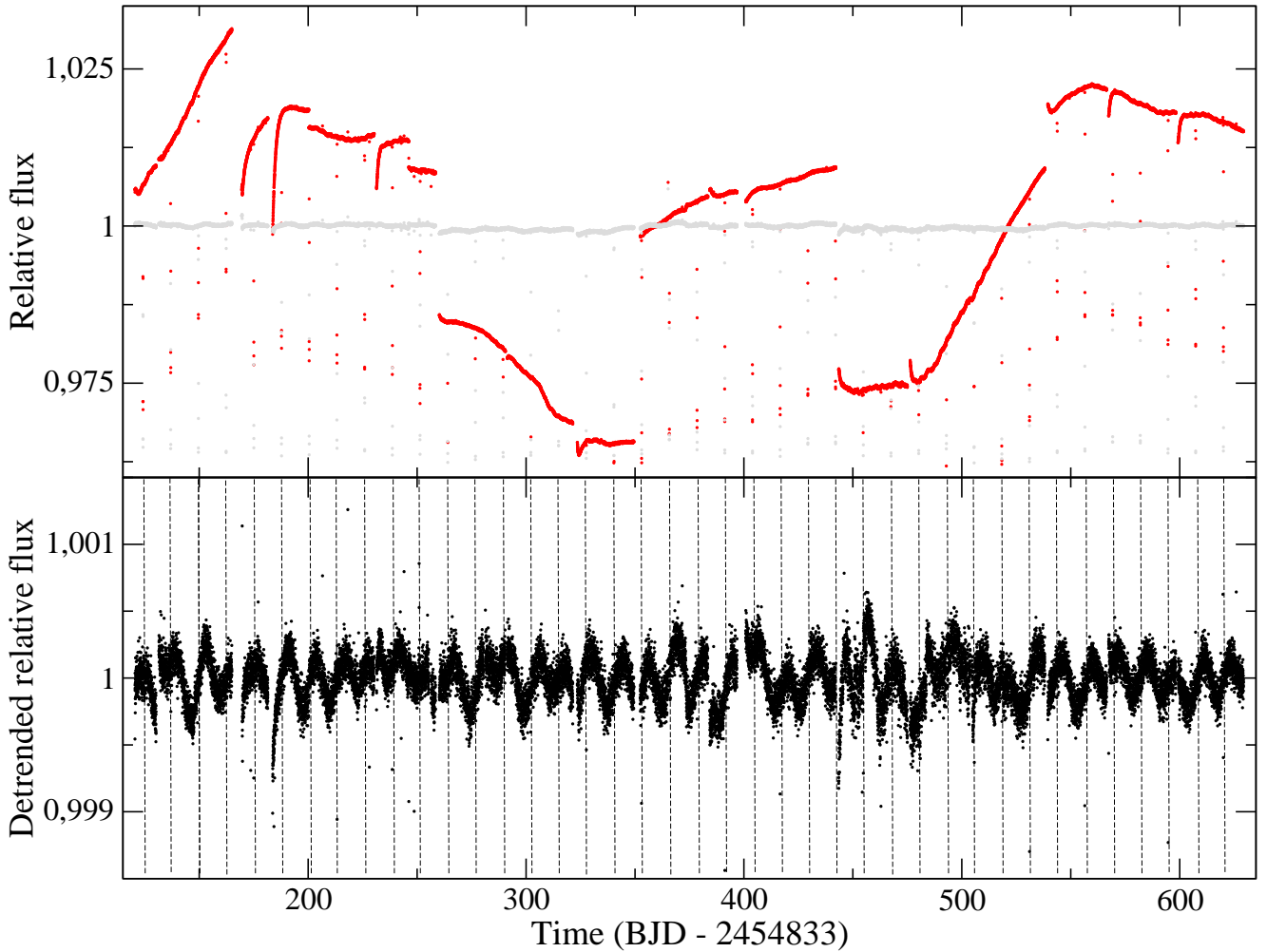}}  % height=9.5cm
\caption{{\it Upper panel:} Kepler SAP light curve of LHS 6343, covering the first six quarters (red dots), and after correction from main trends using cotrending basis vectors (grey dots). {\it Lower panel:} LHS 6343 light curve after low-order polynomial filtering, as adopted for subsequent analysis. The times of mid-transits of LHS 6343 C are marked by vertical dashed lines.
}
\label{lcraw}
\end{figure*}
%%%%%%%%%%%%%%%%%%%%%%%%%%%%%%%%%%%%%%%%%%%%%%%%%%%%%%%%%%%%%

LHS 6343 (KIC 010002261) has been observed by Kepler during all the first six quarters. The time series contains a total of 22976 data points with  $\sim$30 minute cadence and a mean relative precision of $7 \times 10^{-5}$, ranging from May 2009 to September 2010. Both A and B components of the system, separated by 0.\arcsec55 \citep{2011ApJ...730...79J}, are contained inside the Kepler photometric mask used to measure the flux from the target. Its aperture is optimized by the Kepler pipeline for each individual target. Note that since the image scale is 3.\arcsec98 per pixel it is not possible to measure separately the fluxes coming from the A and B components even redefining the photometric mask.

Cotrending basis vectors were applied to the raw data using PyKE pipeline reduction software \footnote{http://keplergo.arc.nasa.gov/PyKE.shtml} to correct systematic trends, which are mainly  related to the pointing jitter of the satellite, detector instabilities and environment variations \citep{Murphy12}. These are optimized tasks to reduce Kepler Simple Aperture Photometry (hereafter SAP) data\footnote{The SAP light curve is a pixel summation time-series of all calibrated flux falling within the optimal aperture.} as they account for the position of the specific target on the detector plane to correct for systematics.  From two to four vectors were used for each quarter to remove the main trends from the raw data (see Fig.~\ref{lcraw}). A low-order ($\leq 4 $) polynomial filtering was then applied to the resulting data for each quarter because some residual trends still remained followed by discontinuities between quarters. These are due to the change of the target position on the focal plane following  each re-orientation of the spacecraft at the end of each quater. Raw SAP data and detrended light curves are shown in Fig.~\ref{lcraw}. As a consequence of this data reduction process, the general trends disappeared, as well as  jumps of the photometry between quarters. On the other hand, any possible intrinsic long-term variability of the object ($\geq 50$ days) has also been removed. The use of low-order  polynomials to detrend each quarter series of data ensures that the frequency and amplitude of any variability with a timescale comparable with the companion orbital period is preserved. The presence of a number of gaps in the data prevents us from using other detrending methods such as Fourier filtering.

The out-of-transit light curve was cleaned by applying a moving box-car median $3\sigma$-clipping filter to identify and remove outliers. Then it was rebinned with a regular 120 min sampling to reduce the computation load for the spot modelling. This is allowed because variations related to spot activity have timescales of the order of one day or longer. The average standard error of the binned observations is $\sim 5.0 \times 10^{-5}$ in relative flux units. Transits were removed assuming the ephemeris of \cite{2011ApJ...730...79J}. The resulting light curve consists of 5716 data points ranging from BJD 2454953.569 to BJD 2455462.163, i.e., with a duration of 508.593 days.

\begin{figure}[]
\centerline{
\includegraphics[width=8.5cm,height=6.3cm]{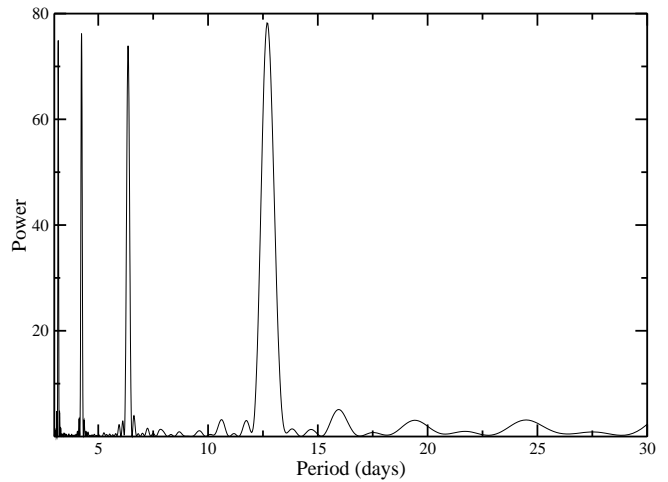}} % {period.eps}} 
\caption{Lomb-Scargle periodogram of the out-of-transit light modulation of LHS 6343. A Gaussian fit was applied to the highest peak to obtain the period of the main modulation and its uncertainty ($12.71\pm0.28$ days).  Peaks corresponding to the harmonics of the main periodicity also appear at higher frequences.
}
\label{fig_period}
\end{figure}

\begin{figure}[]
\centerline{
\includegraphics[width=8.5cm]{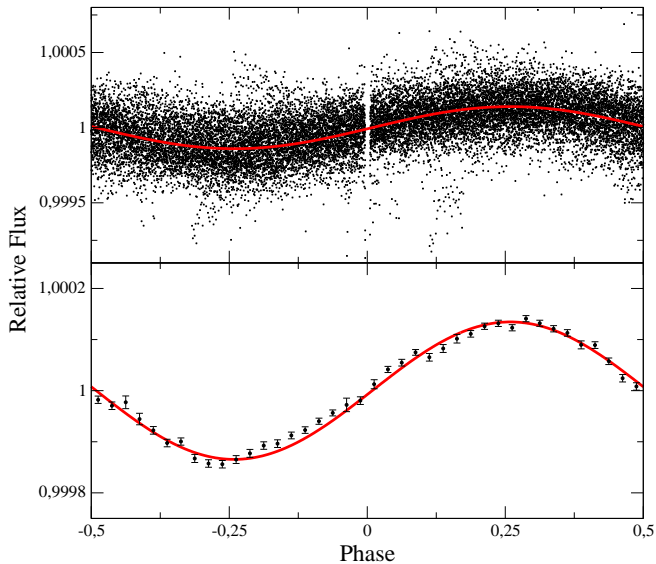}} % {lc_phase.eps}} 
\caption{{\it Upper panel:} Kepler light curve of LHS 6343 phased with a period of 12.71 days (black dots). The transits have been removed leaving a gap centred on phase $0$. The  sinusoidal best fit applied to calculate the phase lead ($\sim98^\circ$) is plotted as a red solid line. {\it Lower panel:} The vertical scale is zoomed in and the Kepler data are plotted in 0.025 phase bins to highlight the deviations from a sinusoidal shape which cause the appearance of harmonics of the main frequency in the periodogram of Fig.~\ref{fig_period}. Error bars are 1-$\sigma$ deviations of the binned flux values. 
}
\label{lc_phase}
\end{figure}

Studying the photometric variability of LHS 6343, we note that the light curve exhibits a periodic modulation with a  low amplitude ($\sim10^{-3}$ in relative flux units, resulting in a signal-to-noise ratio around $15-20$) that remains stable throughout all the time series. The minima of the modulation are separated by a nearly constant phase interval from the transits of the BD companion, as can be seen in the lower panel of Fig.~\ref{lcraw}, which also shows the long-term coherence of the modulation. This suggests that component A is responsible for the observed variability. A first analysis of the out-of-transit light curve was performed through a Lomb-Scargle periodogram, finding only one significant period of $12.71\pm0.28$ days, thus confirming the synchronicity with the LHS~6343~C orbital period (see Fig.~\ref{fig_period}). The other peaks seen in the periodogram correspond to the harmonics of the main periodicity and have heights similar to that of the main peak  because the signal is quasi-periodic, but not exactly sinusoidal in shape. A sinusoidal fit to the light curve with the above  period was performed  to estimate the phase lag, showing that the light minima precede the mid-transits by  $98^{\circ}$  in phase. The phased light curve obtained with all the Q0 - Q6 data is presented in Fig.\ref{lc_phase} with overplotted a sinusoidal best fit, showing the difference in phase between the minima of the photometric modulation and the transits. 

%%%%%%%%%%%%%%%%%%%%%%%%%%%%%%%%%%%%%%%%%%%%%%%%%%%%%%%%%%%%%%%%%
%%%%%%%%%%%%%%%%%%%%%%%%%%%%%%%%%%%%%%%%%%%%%%%%%%%%%%%%%%%%%%%%%
%%%%%%%%%%%%%%%%%%%%%%%%%%%%%%%%%%%%%%%%%%%%%%%%%%%%%%%%%%%%%%%%%
%%%%%%%%%%%%%%%%%%%%%%%%%%%%%%%%%%%%%%%%%%%%%%%%%%%%%%%%%%%%%%%%%

For further analysis and spot modelling, we assume that all the observed variability is caused by LHS 6343 A. This is also supported by the fact that the system is estimated to be  older than 1-2 Gyr from the very low observed chromospheric activity, and most likely near 5~Gyr from brown dwarf mass and radius relations \citep{2011ApJ...730...79J,Baraffeetal98}. Therefore, the  rotation period of the isolated component LHS 6343 B is expected to be longer  than that of the observed modulation
\citep{Barnes03,Barnes07}. 

Since both A and B components of the system fall within the Kepler photometric aperture, it is necessary to correct for the flux contribution from star B before performing any modelling on the light curve. \cite{2011ApJ...730...79J} use the measured $B-V$ colours and $V$ magnitudes  to estimate the relative magnitudes of the stars in the Kepler bandpass, obtaining $\Delta K_{P}=0.74 \pm 0.10$. This yields a flux ratio of $1.97 \pm 0.19$ between star A and B. By assuming that all the modulation comes from star A, we use this value to correct for the dilution of the flux produced by  the component B.
 This hypothesis is valid if the amplitude  $\Delta F_{\rm B}$ of the variability of component B satisfies the relationship: 
\begin{equation}
\frac{\Delta F_{\rm B}}{F_{\rm A}+F_{\rm B}}\ll \frac{\Delta F_{\rm A}}{F_{\rm A}+F_{\rm B}},
\label{fluxr}
\end{equation}
where $\Delta F_{\rm A}$ is the variability of component A and $F_{\rm A} + F_{\rm B}$ the total flux of the system. Given the above flux ratio, this implies:
\begin{equation}
\frac{\Delta F_{\rm B}}{F_{\rm B}} \ll  (1.97\pm0.19)\frac{\Delta F_{\rm A}}{F_{\rm A}} 
\label{sigr}
\end{equation}
In other words, the relative amplitude of the  variability of LHS~6343~B must be much lower than twice that of the A component to ensure that our analysis is consistent. This seems to be reasonable in view of the motivations given above.

\section{Light curve modelling}
\label{sec_model}

\subsection{Transit modelling}
\label{sub_transit}

Although \cite{2011ApJ...730...79J} present a transit light curve analysis, recomputating the system parameters considering the currently available dataset is justified, since the more extended Q0-Q6 transit photometry reveals several features related to spot occultations that would cause a bias in the computation of the transit depth if not properly accounted for. Specifically, when the transiting BD occults a dark feature on the disc of the A component, a relative flux increase appears along the light curve leading to a reduced relative depth of the transit \citep[cf., e.g., ][]{Ballerinietal12}.  

For modelling the transits, we retained the original Kepler time cadence of $\sim 29.4$ minutes without applying the $3\sigma$-clipping filter to the time series. Our fitting procedure for the phased transit photometry is based on the \cite{Mandel2002} analytic model and is performed by using the {\it Transit Analysis Package} \citep{Gazaketal12}, a graphic user-interface IDL tool. This combines a Markov chain Monte Carlo (MCMC) technique to fit light curves and a wavelet-based likelihood function \citep{Carter09}, which is used to compute parameter uncertainties.  The model light curve is resampled and rebinned according to the Kepler long-cadence ($\sim29.4$ minutes) data sampling to avoid systematic deviations, in particular during the ingress and egress phases \citep{Kipping2010}.  The orbital period is fixed to 12.71382 days \citep{2011ApJ...730...79J}, and the adjusted parameters are the inclination $i$, the ratio of the companion and primary radii $R_{\rm C}/R_{\rm A}$, the scaled semimajor axis ($a/R_{\rm A}$), and the time of mid-transit $T_{0}$. Limb darkening is treated by adopting a quadratic approximation, and the coefficients are interpolated for the specific LHS 6343 A stellar parameters from the tabulated values of \cite{Claret11}. An attempt to derive the limb-darkening coefficients from the data fitting while fixing the rest of the parameters has resulted in unphysical values, probably due to a complex brightness profile and the presence of active regions. Both $u_{1}$ and $u_{2}$ have therefore been fixed to the values obtained from the tables. In most cases, the low impact of the limb darkening on the shape of the transit -- note its almost flat-bottomed profile  -- and the existing degeneracy with other parameters, such as the inclination, makes this a better approach than obtaining the coefficients from transit modelling.

The eccentricity and argument of periastron are given by \cite{2011ApJ...730...79J}. A low value of the eccentricity ($e\simeq0.05$) may still affect the transit modelling by increasing its duration by $\sim5\%$ when the argument of periastron is changed from $\omega=0^{\circ}$ to $\omega=180^{\circ}$ \citep[cf. ][]{Perryman11}. Indeed, there is a certain probability that the eccentricity is zero because the value is within two standard deviations from zero
(assuming a Gaussian distribution of the parameter).  Moreover, the eccentricity is a definite positive quantity so any orbital fit to a radial velocity time series with some degree of noise tends to give a non-zero value for this parameter even if the orbit is perfectly circular 
\citep[cf., e.g., ][]{Husnooetal12}.
 The eccentricity is expected to be zero in our system, which is close to synchronization, but a possible excitation by perturbations caused by the distant B companion cannot be excluded in the absence of a specific study. However, a transit analysis with the assumption of a circular orbit was also performed and obtained no significant change in the parameter values. 

%%%%%%%%%%%%%%%%%%%%%%%%%%%%%%%%%%%%%%%%%%%%%%%%%%%%%%%%%%%%%%%%%
\begin{figure}[]
\centerline{
\includegraphics[height=9.5cm,angle=-90]{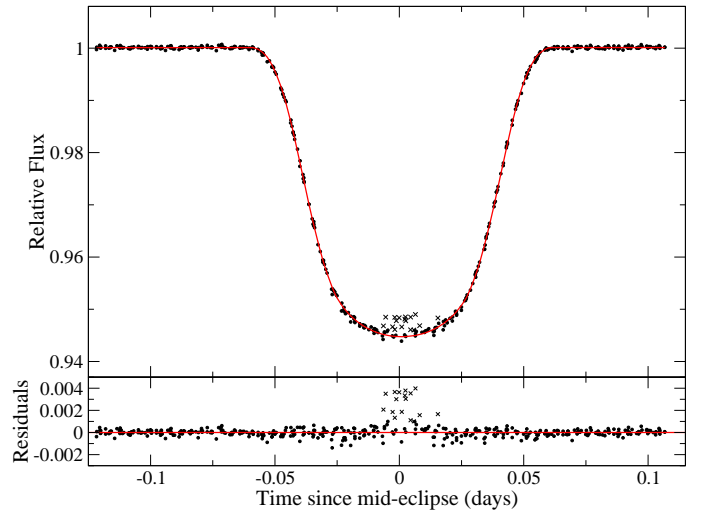}} % {LHS_tapfit.ps}}
\caption{{\it Upper panel:} Kepler LHS 6343 transit photometry from Q0-Q6 (black dots) and the best fit model obtained as described in Sect.~\ref{sub_transit} (red line); crosses indicate data corresponding to spot occultations during the transits. They are not used for modelling the transit. {\it Lower panel:} The residuals of the transit best fit (black dots) and the 
points corresponding to spot occultations (crosses) vs. the time elapsed from mid-transit. }
\label{fig_eclip}
\end{figure}
%%%%%%%%%%%%%%%%%%%%%%%%%%%%%%%%%%%%%%%%%%%%%%%%%%%%%%%%%%%%%%%%%
\begin{table}
\noindent 
\caption{Parameters obtained from the transit light curve modelling of LHS~6343~A.}
\begin{center}
\begin{tabular}{lr}
\hline
\hline
Parameter & Value$^{a}$\\
\hline
Radius ratio ($R_{C}/R_{A}$)          & 0.221 $\pm$ 0.002              \\
Semimajor axis ($a_{R}=a/R_{A}$)      & 45.5 $\pm$ 0.4                 \\
Orbit inclination ($i$)               & 89.60 $\pm$ 0.04               \\
Impact parameter ($b$)                & 0.30 $\pm$ 0.03                \\
Mid transit (BJD-2454957.0)           & 0.2174 $\pm$ 0.0001            \\
Linear limb darkening ($u_{1,A}$)     & 0.457$^{b}$                    \\
Quadratic limb darkening ($u_{2,A}$)  & 0.374$^{b}$                    \\
Eccentricity                          & 0.056 $\pm$ 0.032$^{c}$        \\
Argument of periastron, $\omega$ (deg) & -23 $\pm$ 56$^{c}$            \\
Mean stellar density, $\rho_{A}$ ($\rho_{\odot}$) & 6.8 $\pm$ 0.4            \\
\hline
\label{tab_transit}
\end{tabular}
\end{center}
$^{a}$ Uncertainties correspond to 85.1\% confidence interval from MCMC fitting plus the contribution of systematic errors.

$^{b}$ Fixed values obtained from \cite{Claret11}.

$^{c}$ Fixed values obtained from \cite{2011ApJ...730...79J}.
\end{table}
%%%%%%%%%%%%%%%%%%%%%%%%%%%%%%%%%%%%%%%%%%%%%%%%%%%%%%%%%%%%

\begin{figure*}[]
\centerline{
\includegraphics[width=14cm]{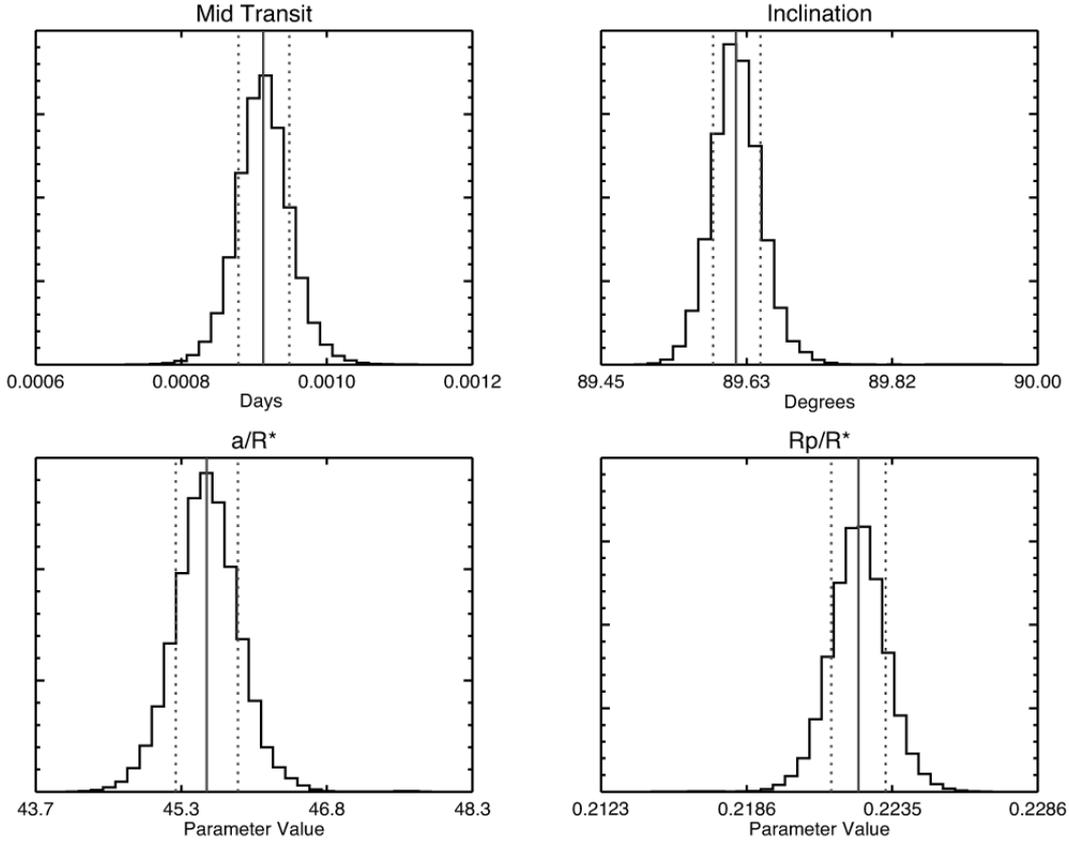}} % {TAP_fit_hist.eps}} 
\caption{Probability density functions for the transit parameters obtained with the MCMC algorithm. The dotted vertical lines correspond to $2-\sigma$ confidence levels.
}
\label{fit_hist}
\end{figure*}

A first modelling of the transit was performed by using a Levenberg-Marquardt algorithm, and  a three-sigma clipping filter was applied to the positive residuals of this fit to identify and remove the data points affected by occulted spots. Then, an MCMC fitting algorithm was applied and the resulting posterior parameter probability distribution was obtained from the combination of ten chains. Several extra runs were performed by fixing each of the parameters in turn to check the consistency of the results and find possible systematic errors that were added to the final uncertainties. The error on the stellar flux ratio (see Sect.~\ref{sec_obs}) was also accounted for in the presented uncertainties. The presence of surface brightness inhomogeneities, associated with stellar activity, makes it especially difficult to  accurately model the transit with our long-cadence data because the induced perturbations are often unresolved.  This produces an increase in the in-transit residuals. The results are presented in Table \ref{tab_transit} and the fit displayed in Fig.~\ref{fig_eclip}. The posterior probability density functions of the parameters obtained by the MCMC  modelling are shown in Fig.~\ref{fit_hist} assuming uniform priors. 

%%%%%%%%%%%%%%%%%%%%%%%%%%%%%%%%%%%%%%%%%%%%%%%%%%%%%%%%%%%%%%%%%
%%%%%%%%%%%%%%%%%%%%%%%%%%%%%%%%%%%%%%%%%%%%%%%%%%%%%%%%%%%%%%%%%

Our mean transit profile, derived from 41 observed transits of LHS~6343~C after removing the spot perturbations, has a better sampling of the ingress and egress intervals allowing us to reduce the uncertainties on the parameters  in comparison with the previous determination by \citet{2011ApJ...730...79J}.  Specifically, we find a sligthly lower value for the relative radius and a higher value for the inclination, although both are within two standard deviations of their previous values, respectively. That we find a very similar result for $R_{C}/R_{A}$ essentially means that spot occultations did not occur or at least were not significant for the five transit events present in the data analysed by \citet{2011ApJ...730...79J}. On the other hand, the results for the relative orbital semimajor axis $a/R_{\rm A}$ are in close agreement. Also the mean stellar density as computed from the transit parameters \citep[e.g., ][]{Seager2003} is consistent with the mass and radius parameters  in Table~\ref{model_param_table}, adopted from \citet{2011ApJ...730...79J}.  

Some systematic trends are observed in the residuals, almost symmetrically placed around the second and third contacts. They could be due to a complex brightness profile along the occulted chord on the stellar disc. In other words, a quadratic limb-darkening law could not be a good approximation for this very late-type star. Short-cadence data should be used to investigate these systematic deviations further along individual transits.

The occultation of the stellar disc by the BD companion during the transits can be used to resolve the fine structure of the spot pattern and study its  evolution \citep[e.g.,][]{SilvaValioetal10,Silva-ValioLanza11}. From the phase-folded data shown in Fig.~\ref{fig_eclip}, we find evidence of  occulted spots that seem to primarily appear close to mid-transit. This could mean that some active regions are associated with the sub-companion point. Therefore, short-cadence data for LHS 6343 would be particularly interesting for further study. This would also help for constraining the projected obliquity and  confirm the quasi-synchronicity of the system \citep[cf., e.g., ][]{Sanchis11a,Sanchis11b}. 

Transit residuals are analysed to look for variability in the transit depth. One of the effects producing such a variability could be a change in the third light contributed by the component B. We find that transit residuals are dominated by noise and do not show any modulation or trend, thus confirming that star B is not active enough to produce a detectable variability and is probably rotating with a much longer period than star A. However, since some scatter in the transit depth may  be caused also by active regions in star~A that are not occulted during transit, it is difficult to use this result to put a strong constraint on the variability of LHS~6343~B.

\subsection{Spot modelling}
\label{sub_spot}

The reconstruction of the surface brightness distribution from the rotational modulation of the stellar flux is an ill-posed problem, because the variation in the flux vs. rotational phase contains information  only on the distribution of the brightness inhomogeneities vs. longitude. The integration over the stellar disc effectively cancels any latitudinal information, particularly when the inclination of the rotation axis along the line of sight is close to $90^{\circ}$, as is assumed in the present case \citep[see Sect.~\ref{sec_params} and ][]{Lanzaetal09a}. Therefore, we need to include a priori information in the light curve inversion process to obtain a unique and stable map. This is  done by computing a maximum entropy (hereafter ME) map, which has been proven to successfully reproduce active region distributions and area variations in the case of the Sun \citep[cf. ][]{Lanzaetal07}. 

Our approach has  been compared with other spot modelling methods by \citet{Mosseretal09} and  \citet{Huberetal10} and with the maps obtained by spot occultations during the planetary transits in the CoRoT-2 system by \citet{Silva-ValioLanza11}.  To model the high-precision space-borne photometry of $\epsilon$~Eri and $k^{1}$~Ceti obtained by  MOST (the Microvariability and Oscillation of STar satellite), \citet{Croll06}, \citet{Crolletal06}, and \citet{Walkeretal07} have developed a sophisticated approach based on a few circular spots. They use simulated annealing and tempering techniques to explore the multi-dimensional model parameter space and Markov chain Monte Carlo methods to assess the parameter uncertainties. The main limitation of these models is that the reduced $\chi^{2}$ values of  their best fits  are significantly greater than the unity. This indicates that the model assumptions, in particular the consideration of only two or three fixed spots, are too simplified to reproduce the true stellar variability down to the level of precision of space-borne photometry.

A similar approach has been applied in the case of the Kepler light curves by \citet{Frascaetal11} and \citet{Frohlichetal12} by considering several spots whose areas evolve with time. 
Although these models allow a convincing determination of the model parameters and their uncertainties, they require a huge amount of computation and their best fits often show systematic deviations suggesting that the spot pattern is more complex than can be described with several circular, non-overlapping starspots. Therefore, we prefer to use a continuos distribution for the spotted area and a regularization approach to find a unique and stable solution.

An important advantage of the ME regularization is that the multi-dimensional parameter space of the objective function to be minimized (cf. Eq.~\ref{eq_gof}) has a smoother landscape than  
the $\chi^{2}$ landscape of a multi-spot model, thanks to the addition of the configurational entropy term \citep[the function $S$ defined below; cf. the analogous problem of image restoration in ][]{BryanSkilling80}. This effectively reduces the problems associated with the presence of several $\chi^{2}$ relative minima falling in widely separated regions of the parameter space \citep[cf. ][]{Crolletal06,Walkeretal07}. Moreover, we effectively average over many degrees of freedom by only considering the distribution of the spotted area vs. the stellar longitude, which is the only information that can be directly extracted from our unidimensional time series. Therefore, the basic role of our regularization is that of smoothing the distribution of the spot filling factor versus longitude, thus reducing the impact of the noise on the light curve inversion process.

In our model, the star is subdivided into 200 surface elements, namely  200  squares of side $18^{\circ}$, with  each element containing unperturbed photosphere, dark spots, and facular areas. The fraction of an element covered by dark spots is indicated by the filling factor $f$,  the fractional  area of the faculae is $Qf$, and the fractional area of the unperturbed photosphere is $1-(Q+1)f$. 
The contribution to the stellar flux coming from the $k$-th surface element at the time $t_{j}$, where $j=1,..., N$   is an index numbering the $N$ points along the light curve, is given by
\begin{eqnarray}
\Delta F_{kj} & = & I_{0}(\mu_{kj}) \left\{ 1-(Q+1)f_{k} + c_{\rm s} f_{k} +  \right. \nonumber \\
  & & \left.  Q f_{k} [1+c_{\rm f} (1 -\mu_{kj})] \right\} A_{k} \mu_{kj} {w}(\mu_{kj}),
\label{delta_flux}
\end{eqnarray}
where $I_{0}$ is the specific intensity in the continuum of the unperturbed photosphere at the isophotal wavelength of the observations; $c_{\rm s}$ and $c_{\rm f}$ are the spot and facular contrasts, respectively \citep[cf. ][]{Lanzaetal04}; $A_{k}$ is the area of the $k$-th surface element,
\begin{equation}
 {w} (\mu_{kj}) = \left\{ \begin{array}{ll} 
                      1  & \mbox{if $\mu_{kj} \geq 0$}  \\
                      0 & \mbox{if $\mu_{kj} < 0$ }
                              \end{array} \right. 
\end{equation}
is its visibility; and 
\begin{equation}
\mu_{kj} \equiv \cos \psi_{kj} = \sin i \sin \theta_{k} \cos [\ell_{k} + \Omega (t_{j}-t_{0})] + \cos i \cos \theta_{k},
\label{mu}
\end{equation}
is the {cosine} of the angle $\psi_{kj}$ between the normal to the surface element and the direction of the observer, with $i$ being the inclination of the stellar rotation axis along the line of sight; $\theta_{k}$ the colatitude and $\ell_{k}$ the longitude of the $k$-th surface element; $\Omega$ {denotes} the angular velocity of rotation of the star ($\Omega \equiv 2 \pi / P_{\rm rot}$, with $P_{\rm rot}$ the stellar rotation period), and $t_{0}$ the initial time. The specific intensity in the continuum varies according to a quadratic limb-darkening law, as adopted by \citet{Lanzaetal03} for the case of the Sun, viz. $I_{0} \propto a_{\rm p} + b_{\rm p} \mu + c_{\rm p} \mu^{2}$. The stellar flux computed at the time $t_{j}$ is then $F(t_{j}) = \sum_{k} \Delta F_{kj}$. To warrant a relative precision of the order of $10^{-5}$ in the computation of the flux $F$, each surface element is further subdivided into $1^{\circ} \times 1^{\circ}$-elements,  and their contributions, calculated according to Eq.~(\ref{delta_flux}), are summed up at each given time to compute the contribution of the $18^{\circ} \times 18^{\circ}$ surface element to which they belong.  

We fit the light curve by varying the value of $f$ over the surface of the star, while $Q$ is held constant. Even fixing the rotation period, the inclination, and the spot and facular contrasts \citep[see ][ for details]{Lanzaetal07}, the model has 200 free parameters and suffers from  non-uniqueness and instability. To find a unique and stable spot map, we apply ME regularization by minimizing a functional $Z$, which is a linear combination of the $\chi^{2}$ and  the entropy functional $S$; i.e.,
\begin{equation}
Z = \chi^{2} ({\vec f}) - \lambda S ({\vec f}),
\label{eq_gof}
\end{equation}
where ${\vec f}$ is the vector of the filling factors of the surface elements, $\lambda > 0$   a Lagrangian multiplier determining the trade-off between light curve fitting and regularization. The expression for $S$ is given in \citet{Lanzaetal98}.  The entropy functional $S$  attains its maximum value when the star is {free of spots}. Therefore, by increasing the Lagrangian multiplier $\lambda$, we increase the weight of $S$ in the model, and the area of the spots  is progressively reduced.
This gives rise to systematically negative residuals between the observations and the best-fit model when
$\lambda > 0$. The optimal value of $\lambda$ depends on the information content of the light curve, which in turn depends on the ratio of the amplitude of its rotational modulation to the average error of its  points. In the case of LHS 6343 A, the amplitude of the  rotational modulation is $\sim 0.001$, while the average error of the  points is $\sim 5 \times 10^{-5}$ in relative flux units, giving a signal-to-noise ratio of $\sim 20 $. The significant impact of the noise on the light curve makes the regularization a critical process, and a different criterion than that applied for CoRoT-2 \citep[see ][]{Lanzaetal09a} or CoRoT-6 \citep[see ][]{Lanzaetal11} must be adopted, because in those cases the signal-to-noise ratio was $>100$. For the present case, we increase the regularization until we obtain $| \mu_{\rm reg} | = \beta \epsilon_{0}$, where $\mu_{\rm reg}$ is the mean of the residuals of the regularized light curve, and $\epsilon_{0} \equiv \sigma_{0}/\sqrt{N}$ is the standard error of the  residuals of the unregularized light curve (i.e., obtained with $\lambda=0$), defined as the ratio of their standard deviation $\sigma_{0}$  to the square root of the number of data points $N$ in each individual segment of duration $\Delta t_{\rm f}$ fitted by our model (see below). The parameter $\beta \geq 1 $ depends on the information content of the light curve and is unity when the signal-to-noise ratio is of the order of 100 or greater. In our case, we determine the optimal value of $\beta$ a posteriori, by increasing $\lambda$  until the small random spots associated with the noise are strongly smoothed, while the deviation of the regularized model from the data points is still acceptable (see Sect.~\ref{sub_reslc} and  Appendix~\ref{appendixa} for the effects  of varying $\beta$ on the solution). 

In the case of the Sun, by assuming a fixed distribution of the filling factor, it is possible to obtain a good fit of the irradiance changes only for a limited time interval $\Delta t_{\rm f}$, not exceeding 14 days, which is the lifetime of the largest sunspot groups dominating the irradiance variation. In the case of other active stars, the value of $\Delta t_{\rm f}$ must be determined from the observations themselves, looking for the longest time interval that allows  a good fit with the applied model (see Sect.~\ref{sec_params}). 

The optimal values of the spot and facular contrasts and of the facular-to-spotted area ratio $Q$ in stellar active regions are  unknown a priori. In our model the facular contrast $c_{\rm f}$ and the parameter $Q$ enter as the product $c_{\rm f} Q$, so we can fix $c_{\rm f}$ and vary $Q$, estimating its best value 
 by minimizing the $\chi^{2}$ of the model, as shown in Sect.~\ref{sec_params}. Since there are many free parameters in the ME model is large, for this specific application we make use of the model of \citet{Lanzaetal03}, which fits the light curve by assuming only three active regions to model the rotational modulation of the flux plus a uniformly distributed background to account for the variations in the mean light level. This procedure is the same as used to fix the value of $Q$ in the cases of CoRoT-2, CoRoT-4, CoRoT-7, and CoRoT-6
\citep[cf. ][2010, 2011]{Lanzaetal09a,Lanzaetal09b}.  

We  assume an inclination of the rotation axis of LHS 6343 A of $ i \simeq 90^{\circ}$ in most of our models (see Sect.~\ref{sec_params}), coming from the results of the transit  modelling (see Sect.~\ref{sub_transit}). Such a high inclination implies that negligible information on the spot latitude  can be extracted from the rotational modulation of the flux, so that the ME regularization virtually puts all the spots around the subobserver latitude (i.e., $\approx 0^{\circ}$) to minimize their area and maximize the entropy. Therefore, we are limited to mapping  only the distribution of the active regions vs. longitude, which can be done with a resolution better than   $\sim 50^{\circ}$ \citep[cf. ][]{Lanzaetal07,Lanzaetal09a}. Our ignorance of the true facular contribution  may lead  to systematic errors in the active region longitudes derived by our model because faculae produce an increase in the flux when they are close to the limb, leading to a systematic shift of the longitudes of the active regions used to reproduce the observed flux modulation,   as discussed  by \citet{Lanzaetal07} {for} the case of the Sun and illustrated by \citet[][ cf.~ Figs.~4 and 5]{Lanzaetal09a} for CoRoT-2.

Given the dependence of our modelling approach on several parameters, we present a detailed discussion of their impact on our results in Appendix~\ref{appendixa}.

\subsection{Model parameters}
\label{sec_params}

The fundamental stellar parameters are taken from \cite{2011ApJ...730...79J} and are listed in Table \ref{model_param_table}. A quadratic limb-darkening law is adopted for the stellar photosphere, viz., $I(\mu) \propto a_{\rm p} + b_{\rm p} \mu + c_{\rm p} \mu^{2}$, where the coefficients $a_{\rm p}$, $b_{\rm p}$,  and $c_{\rm p}$ have been obtained for $T_{\rm eff}=3130$~K, log $g=4.85$ $[$cgs$]$, and solar abundances from the theoretical values computed by \cite{Claret11} from Phoenix model atmospheres.

%%%%%%%%%%%%%%%%%%%%%%%%%%%%%%%%%%%%%%%%%%%%%%%%%%%%%%%%%%%%%%%%%%%%%%%%%%%%%%%%
\begin{table}
\noindent 
\caption{Parameters adopted for the light curve  modelling of LHS~6343~A.}
\begin{center}
\begin{tabular}{lrr}
\hline
\hline
Parameter &  & Ref.$^{a}$\\
\hline
Star mass ($M_{\odot}$) & 0.370 & J11  \\
Star radius ($R_{\odot}$) & 0.378 & J11  \\
$T_{\rm eff}$ (K) & 3130 &  J11 \\
$\log g$ (cm s$^{-2}$) & 4.85 & J11 \\ 
$a_{\rm p}$ & 0.1695 & CB11 \\
$b_{\rm p}$ & 1.2041 & CB11 \\
$c_{\rm p}$ & -0.3736 & CB11 \\ 
$P_{\rm rot}$ (days) & 12.7138 & J11 \\
$\epsilon$ & $6.09 \times 10^{-6}$ &  He12\\ 
Inclination (deg) & 89.60 & He12  \\
$c_{\rm s}$  & 0.536 & He12 \\
$c_{\rm f}$  & 0.115 & L04 \\ 
$Q$ & 0.0, 8.0  & He12 \\ 
$\Delta t_{\rm f}$ (days) & 8.769 & He12 \\ 
\hline
\label{model_param_table}
\end{tabular}
\end{center}
$^{a}$ References: CB11:  \citet{Claret11}; He12: present work; J11: \citet{2011ApJ...730...79J}; L04: \citet{Lanzaetal04}.  
\end{table}
%%%%%%%%%%%%%%%%%%%%%%%%%%%%%%%%%%%%%%%%%%%%%%%%%%%%%%%%%%%%%%%%%%%%%%%%%%

The rotation period for modelling the spotted photosphere was fixed to the companion orbital period. This assumption comes from the  periodogram analysis giving $P_{\rm rot}=12.71 \pm 0.28$~days (see Fig. \ref{fig_period}), which indicates that the system is close to synchronization. As we see in Sect. \ref{sub_activ}, any deviation of stellar rotation from synchronization will appear on the ME map as a longitude drift of the active regions versus time. 

The rotation period of 12.71 days is used to compute the polar flattening of the star due to the centrifugal potential in the Roche approximation, yielding a relative difference between the equatorial and the polar radii $\epsilon$ of $6.09 \times 10^{-6}$. The effect of this flattening on the modulation caused by spots is negligible. The relative difference in flux for a $\sim 2 \%$ spot coverage would be less than $10^{-6}$ as a consequence of the gravity darkening in the equatorial regions of the star.

Owing to the lack of a measurement of the Rossiter-McLaughlin effect or precise $v \sin i$ data, the inclination of the stellar axis of LHS 6343 A cannot be  constrained. The limit on $v \sin i < 2$ km~s$^{-1}$ given by \cite{2011ApJ...730...79J} does not help because the above rotation period and the radius yield an equatorial rotation velocity of $\sim 1.5$ km~s$^{-1}$. Nevertheless, we assume that the system is spin-orbit aligned; that is, its inclination is assumed to be equal to that of the orbit of the BD companion, i.e. $89 \fdg 60 \pm 0\, \fdg 04$. The inclination of the stellar spin axis along the line of sight is the only parameter required to characterize the geometry of the stellar rotation for modelling the out-of-transit light modulation. If we had been interested in also modelling the occultations of starspots during transits, the sky-projected angle between the stellar spin and the orbital angular momentum would have been required, too \citep[e.g., ][]{Sanchis-Ojedaetal12}.

The maximum time interval $\Delta t_{\rm f}$ that our model can accurately fit with a fixed distribution of active regions has been determined by means of a series of tests dividing the total interval, $T=508.5936$ days, into $N_{f}$ equal segments, i.e., $\Delta t_{\rm f} = T/N_{\rm f}$. The simple three-spot model of \citet{Lanzaetal03} has been applied to fit the individual intervals to reduce the number of free parameters. The optimal $N_{\rm f}$ is given by the minimum value that allows us a good fit to the light curve as measured by the $\chi^{2}$ statistics. For $N_{\rm f}<58$, the quality of the fit starts to degrade significantly, so we adopt this as the optimal number of segments, giving $\Delta t_{\rm f}=8.769$ days  (cf. Appendix~\ref{appendixa}).  In any case,  $\Delta t_{\rm f}>P_{\rm rot}/2$ is required to adequately sample the light modulation produced by a rotating active region, providing us with information on its longitude and area. On the other hand, a too long $\Delta t_{\rm f}$ would result in a low time resolution of our model, so that starspot evolution and migration would be more difficult to trace.

%%%%%%%%%%%%%%%%%%%%%%%%%%%%%%%%%%%%%%%%%%%%%%%%%%%%%%%%%%%%%%%%%
\begin{figure}[]
\centerline{
\includegraphics[width=8.5cm,height=5.8cm]{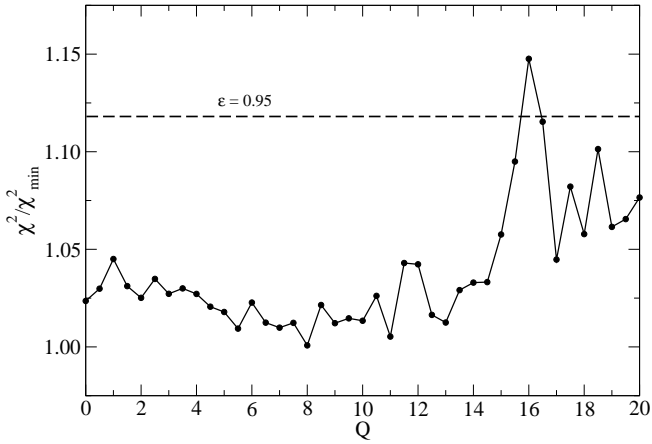}} % {qchi.eps}} 
\caption{The ratio of the $\chi^{2}$ of the composite best fit of the entire time series  to its  minimum value vs. the parameter $Q$, i.e., the ratio of the facular area to the cool spot area in active regions. The horizontal dashed line indicates the $95\%$ confidence level for $\chi^{2}$/$\chi^{2}_{\rm min}$. This does not put any significant constraint on the interval of acceptable $Q$ values owing to the low signal-to-noise ratio of the data.
}
\label{fig_qchi}
\end{figure}
%%%%%%%%%%%%%%%%%%%%%%%%%%%%%%%%%%%%%%%%%%%%%%%%%%%%%%%%%%%%%%%%%

The spot temperature contrast is estimated by the correlation found in Fig.~7 of  \citet{Berdyugina05} that shows that the contrast decreases with decreasing stellar effective temperature.  Thus, we adopt  a spot temperature deficit of $\Delta T_{\rm s}=-250$ K for LHS 6343 A and compute the spot's intensity contrast considering blackbody emission for the spots and the photosphere in the Kepler passband, yielding $c_{\rm s}=0.536$. This is not a critical parameter because a variation in $c_{\rm s}$ will only cause a change in the overall spot filling factor, but  will not affect either spot longitudes or their time evolution, as discussed in detail by \citet{Lanzaetal09a}. The facular contrast is assumed to be solar-like with $c_f=0.115$ \citep{Lanzaetal04}.

The  ratio $Q$ between the facular and the spotted areas in active regions is estimated by means of the three-spot model introduced by \citet{Lanzaetal03} simultaneously with the optimal value of  $\Delta t_{\rm f}$. Both the flux increase due to the facular component when an active region is close to the limb and the flux decrease due to its spots when that region transits across the central meridian of the disc are modelled by this approach,  allowing us to derive the optimal value of $Q$, if the signal-to-noise ratio of the data is high enough. 

In Fig.~\ref{fig_qchi} we plot the ratio $\chi^{2}/ \chi^{2}_{\rm min}$ of the total $\chi^{2}$ of the composite best fit of the entire time series to its minimum value $\chi^{2}_{\rm min}$, versus $Q$. Given the low signal-to-noise ratio of the photometry in our case, we find that there is no clear $\chi^{2}$-minimum vs. $Q$ so its value is  poorly constrained with the best solutions falling in the range between $\sim 5$ and $\sim 11$. Therefore, we adopt $Q=8.0$ for further analysis. This  facular-to-spotted area ratio is similar to the one found in the case of the Sun \citep[$Q=9$, cf. ][]{Lanzaetal07}.
%Moreover, this value of $Q$ together with the adopted facular contrast $c_{\rm f}=0.115$ implies a limb $\sim1.8$ brighter than without faculae , which is in rough agreement with the finding from the transit residuals (see Sect.~\ref{sub_transit}).

%%%%%%%%%%%%%%%%%%%%%%%%%%%%%%%%%%%%%%%%%%%%%%%%%%%%%%%%%%%%%%%%%
\begin{figure*}[]
\centerline{
\includegraphics[width=18cm,height=9cm]{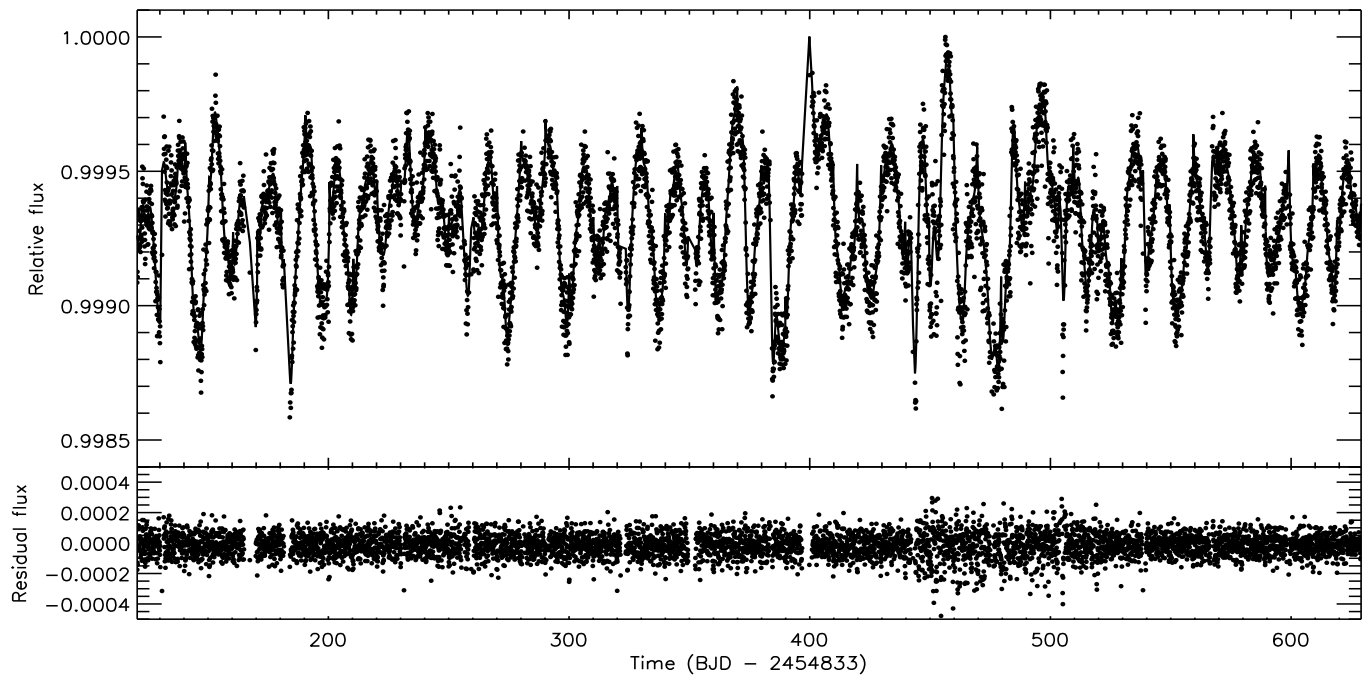}} % {gen-fit.ps}} 
\caption{{\it Upper panel:} The out-of-transit light curve of LHS~6343~A (dots) and its  ME-regularized  best fit  for a facular-to-spotted area ratio $Q=8.0$ (solid line) during the time interval between BJD~2454953.569 and BJD~2455462.163. The flux is normalized to the maximum observed flux along the time series.  {\it Lower panel:} The corresponding residuals.
}
\label{lc_bestfit}
\end{figure*}
%%%%%%%%%%%%%%%%%%%%%%%%%%%%%%%%%%%%%%%%%%%%%%%%%%%%%%%%%%%%%%%%%

\section{Results}
\label{sec_results}

\subsection{Light curve model}
\label{sub_reslc}

The composite best fit obtained with the ME regularization and $Q=8.0$ is shown in Fig.~\ref{lc_bestfit} (upper panel) together with the residuals (lower panel). From a best fit without regularization  ($\lambda=0$)  applied to the whole time series  (Q0$-$Q6), we obtain a mean $\mu_{\rm res}=8.308 \times 10^{-8}$ and a standard deviation of the residuals $\sigma_{0}=6.778 \times 10^{-5}$ in relative flux units, which is slightly higher than the standard error of the mean points. This can be due to short time scale fluctuations that cannot be fitted with our approach that  reproduces variations on a timescale comparable to $\Delta t_{\rm f}$ or longer. On average, we have $N \sim 98$ data points per fitted subset $\Delta t_{\rm f}$.

The Lagrangian multipliers $\lambda$ for the regularized ME models are iteratively adjusted until the mean of the residuals $\mu_{\rm reg}=-1.365 \times 10^{-5} \simeq -\beta (\sigma_{0}/\sqrt{N})$, where $\beta=2$ is established as the optimal value that still leads to an acceptable fit while reducing the information content of the spot maps as much as possible, i.e., maximizing the entropy. A lower value of $\beta$ improves the fit slightly over small sections of the light curve, but introduces several small spots that change from one time interval to the next without any clear regularity. On the other hand, by selecting $\beta =2$, we obtain smoother maps with concentrated groups of spots that evolve regularly from one $\Delta t_{\rm f}$ interval to the next, which indicates that we are properly modelling the overall distribution of the active regions on the star. A higher level of regularization ($\beta>2$) gives unacceptably larger, systematically negative residuals because the maximum entropy criterion reduces the spotted area too much. At the same time, it concentrates the spots into  isolated groups, and a pattern with three active regions equally spaced in longitude (i.e, with $\approx 120^{\circ}$ gaps) begins to appear. This pattern still shows the general migration  of the multiple spots appearing in the  maps with the optimal regularization (see Sect.~\ref{sub_activ}), but it is clearly an artefact of the over-regularization, due to   the limited signal-to-noise level of our data that  favours a reproduction of the light variations by a decreasing number of spots when the regularization is increased (cf. Appendix~\ref{appendixa}).

In either case, the low signal-to-noise ratio of our data  requires a relatively strong regularization, and as a consequence, misfits sometimes appear at the matching points between time intervals. Therefore, by decreasing 
$\Delta t_{\rm f}$, the overall quality of the best fit is not increased in our case. Nevertheless, looking at the residuals of our regularized composite best fit, we see a generally good reproduction of the light variations 
without systematics that could be attributed to sizable active regions evolving on timescales significantly shorter than $\Delta t_{\rm f}$. We notice that the noise level increases after BJD~2455276 (corresponding to BJD$-2454833=443$ in Fig.~\ref{lc_bestfit}), which coincides with the beginning of quarter Q5 and returns to the previous level after BJD~2455371 (BJD$-2454833=538$), i.e., at the beginning of quarter Q6. This is a consequence of the rotation of the focal plane of the Kepler telescope between successive quarters, which makes the target be measured on a different~CCD.

%%%%%%%%%%%%%%%%%%%%%%%%%%%%%%%%%%%%%%%%%%%%%%%%%%%%%%%%%%%%%%%%%
\begin{figure}[]
\centerline{
\includegraphics[height=9.5cm,angle=90]{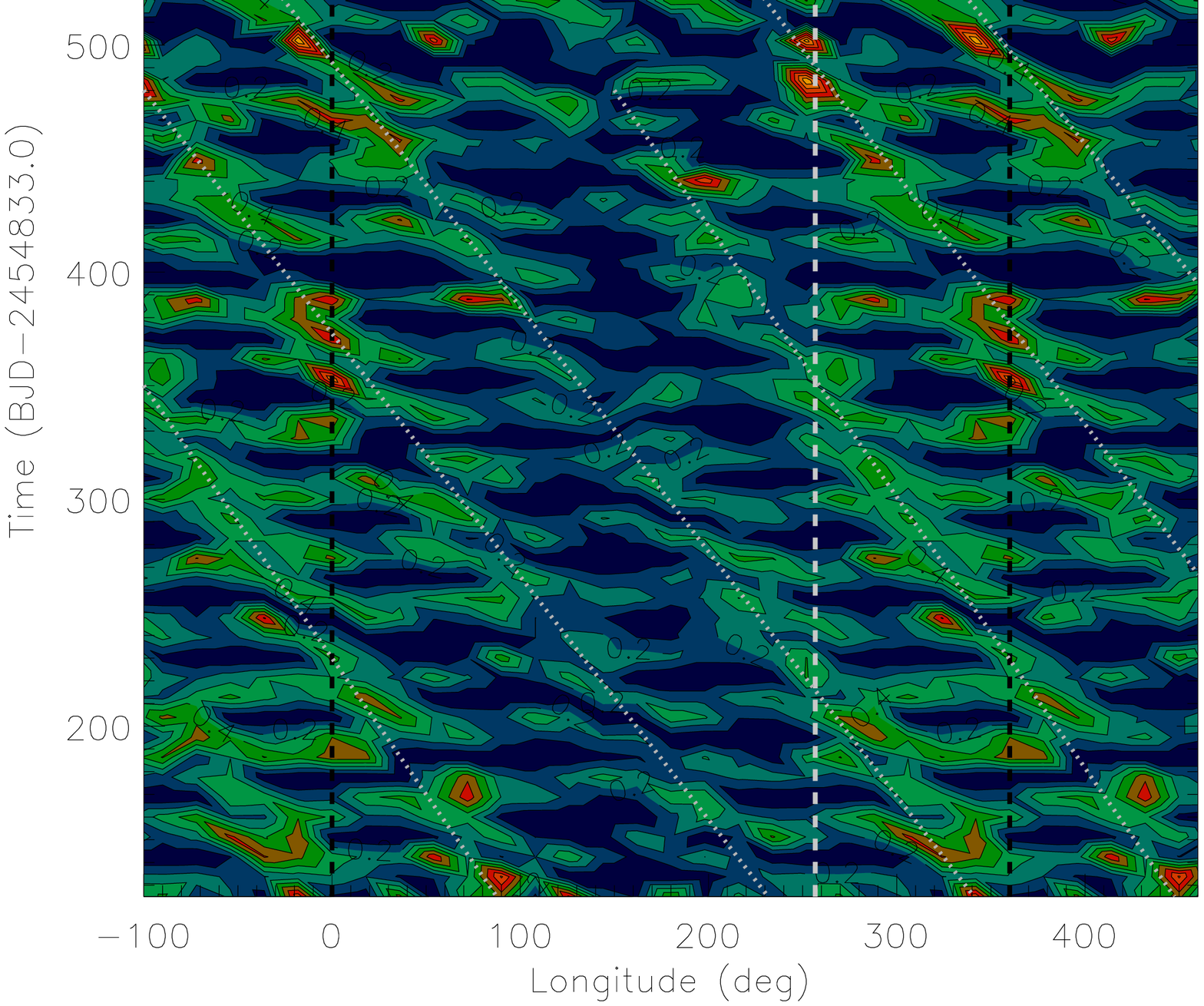}}  % {spot_contours2_sdr.ps} 
\caption{Isocontours of the ratio $f/f_{\rm max}$, where $f$ is the spot covering factor and $f_{\rm max}= 0.00059$ its maximum value,  versus time and longitude for the ME models with $Q=8.0$. The two dashed black lines mark longitudes $0^{\circ}$ and $360^{\circ}$ beyond which the distributions are repeated to easily follow spot migration. The dashed white line marks the longitude of the subcompanion point, which is fixed in this reference frame at $\sim257^{\circ}$. The contour levels are separated by $\Delta f= 0.1 f_{\rm max}$ with yellow and orange indicating the maximum covering factor and dark blue the minimum. The dotted white lines trace the migration of the active regions associated with each active longitude (see the text). }
\label{spot_cont}
\end{figure}
%%%%%%%%%%%%%%%%%%%%%%%%%%%%%%%%%%%%%%%%%%%%%%%%%%%%%%%%%%%%%%%%%

\subsection{Distribution of active regions and stellar rotation}
\label{sub_activ}
 
The distribution of the spot filling factor $f$ versus longitude and time is plotted in Fig. \ref{spot_cont} for our ME model with $Q=8.0$ adopting a reference frame rotating with the orbital period of the BD companion, i.e., $12.7138$~days. The origin of the longitude frame corresponds to the subobserver point at the beginning of the Kepler time series (i.e. BJD 2454953.569) and the longitude increases in the same direction as the stellar rotation and the orbital motion of the companion. The subcompanion longitude is fixed at $\sim 257^{\circ}$ in this reference frame.

Several individual active regions can be identified from the map in Fig. \ref{spot_cont} showing lifetimes around $15-20$ days.  Three  dotted lines have been drawn  connecting the longitudes with higher filling factors 
to trace the overall migration of the spots. They identify three long-lived active longitudes that 
  migrate backwards at the same constant rate.  This indicates that the rotation of the star is somewhat slower than that of our reference frame following the orbital motion of the BD companion.
 The similarity of the migration rates among the three active longitudes suggests that either the spot latitudes are confined to a thin belt or the star has no detectable differential rotation. 
 
By considering the longitudes of the relative maxima of the spot covering factor in Fig. \ref{spot_cont}, a linear fit is performed to find the relative rotation rates associated with the steady migration of the three main starspots.  The results are listed in Table \ref{tab_periods} for two models having a different facular-to-spotted area ratio $Q$. By assuming a significant facular contribution ($Q=8.0$), a mean rotation period of $13.137 \pm 0.011$~days is obtained. This is a significantly slower rotation than the companion orbital motion, showing that the system is slightly out of synchronism. The stellar rotation, as traced by starspots, appears to be rigid since there is no significant difference between the highest and the lowest migration rates. The migration rates for a spot model assuming only dark spots ($Q=0$) are also reported in Table \ref{tab_periods}, because the longitudes of the spots derived from the maximum entropy modelling may depend on the value of $Q$  \citep{Lanzaetal07,Lanzaetal09a}. Nevertheless, very similar results are obtained, thus confirming that the presence of faculae is not actually required to adequately fit  the low signal-to-noise photometry of LHS~6343~A  (cf. Sect.~\ref{sec_params}). In the case of no faculae, a virtually identical mean rotation period of  $13.129 \pm 0.015$~days is obtained from the migration of the spots.

%%%%%%%%%%%%%%%%%%%%%%%%%%%%%%%%%%%%%%%%%%%%%%%%%%%%%%%%%%%%%%%%%
\begin{table}
\noindent 
\caption{Relative migration rates $\Delta\Omega/\Omega$ obtained from linear fits to the migration of the main active longitudes in Fig. \ref{spot_cont}, and the corresponding mean rotation period $P_{\rm rot}$.}
\begin{center}
\tiny
\begin{tabular}{ccccc}
\hline
\hline
$Q$ & AL$_{1}$ & AL$_{2}$ & AL$_{3}$ & $P_{\rm rot}$\\
 & ($10^{-2}$ deg/d) & ($10^{-2}$ deg/d) & ($10^{-2}$ deg/d) & (d) \\
\hline
0.0 & $ -3.20 \pm 0.20$ & $ -3.34 \pm 0.11$ & $-3.24 \pm 0.17$ & $13.129 \pm 0.015$ \\
8.0 & $ -3.34 \pm 0.15$ & $-3.34 \pm 0.14$ & $-3.30 \pm 0.20$ & $13.137 \pm 0.011$\\
\hline
\label{tab_periods}
\end{tabular}
\end{center}
\end{table}
%%%%%%%%%%%%%%%%%%%%%%%%%%%%%%%%%%%%%%%%%%%%%%%%%%%%%%%%%%%%%%%%%%%%%%%%%%

 An intriguing phenomenon seen in Fig.~\ref{spot_cont} is that the spot filling factor shows a persistent relative minimum in the longitude range between $\approx 100^{\circ}$ and $\approx 250^{\circ}$, while it has a relative maximum  between $\approx 250^{\circ}$ and $\approx 450^{\circ}$.  In other words, the  filling factor appears to be systematically enhanced when the migrating starspots approach and cross a longitude leading the subcompanion longitude by $\sim 100^{\circ}$  in our reference frame.
Since the passage of that longitude across the centre of the stellar disc precedes that of the BD companion, this agrees with the light curve that shows a relative light minimum $\sim 3.5$~days before each transit. This modulation has persisted for more than 500 days, as far as the Kepler data currently cover. 

\begin{figure}[]
\centerline{
\includegraphics[height=9cm,angle=90]{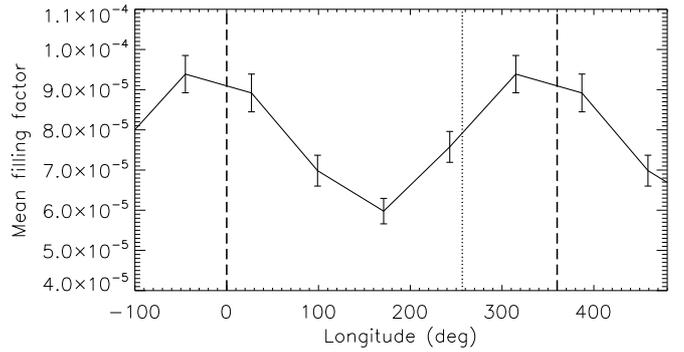}}  % {spot_long.ps} 
\caption{The spotted area in $72^{\circ}$ longitude bins averaged versus time as derived from the ME regularized models with $Q=8.0$. The error bars indicate the standard error of the spotted area in the corresponding bin. The dotted line marks the subcompanion point longitude, while the dashed lines indicate the longitudes $0^{\circ}$ and $360^{\circ}$ beyond which the distribution is repeated for clarity.
}
\label{spot_long}
\end{figure}

This enhancement of the spot filling factor is highlighted in Fig.~\ref{spot_long}, where the spot coverage factor integrated over the time is plotted vs. longitude in bins of 72$^{\circ}$. A wide maximum is evident around $\sim350^{\circ}$, meaning that activity tends to be concentrated in a region preceding the subcompanion longitude. This agrees closely with what can be seen in Fig.~\ref{lcraw}, and especially in the phase-folded light curve (Fig.~\ref{lc_phase}). Moreover, this explains the results obtained by the Lomb-Scargle periodogram (Fig.~\ref{fig_period}) because only the frequency of the main modulation associated with this activity enhancement, phased to the companion C, is detected by that method. On the other hand, ME modelling allows us to map individual starspots  and follow their migration across the longitude where they show an  enhancement. Unfortunately, no information on the spot latitudes can be derived with our models, so we cannot establish whether the enhancement of activity at this longitude is produced in a single latitude or if it covers a range of latitudes. As discussed in Sect.~\ref{sec_discus}, these results strongly suggest some magnetic interaction between LHS~6343~A and its BD companion LHS~6343~C, thus causing spots to mainly appear at a fixed longitude in the reference frame of the companion orbital motion.

%%%%%%%%%%%%%%%%%%%%%%%%%%%%%%%%%%%%%%%%%%%%%%%%%%%%%%%%%%%%%%%%%
%%%%%%%%%%%%%%%%%%%%%%%%%%%%%%%%%%%%%%%%%%%%%%%%%%%%%%

\begin{figure}[]
\centerline{
\includegraphics[height=9cm,angle=90]{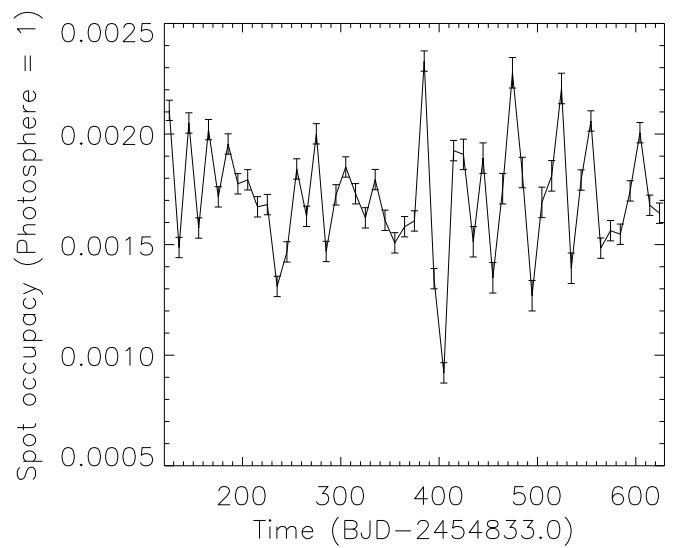}}  % {spot_area.ps};  height=9 cm
\caption{The total spotted area versus time as derived from the regularized ME analysis for $Q=8.0$. The 
error bars correspond to three standard deviations.}
\label{spot_area}
\end{figure}

The variation in the total spotted area vs. the time is shown in Fig.~\ref{spot_area}. Only slight variations can be seen with a typical timescale of 20-30 days, which can be associated with the typical spot lifetime. Short-term oscillations are most likely related to the noise present in our time series, especially  during quarter Q5, residual trends, or gaps in the data (e.g., the one near day $\sim400$, see Fig.~\ref{lcraw}). Long-term trends have been removed when detrending the time series from Kepler instrumental effects (see Sect.~\ref{sec_obs}), so the mean value of the total spotted area stays approximately constant.  The absolute value of the area depends on the adopted spot contrast $c_{\rm s}$ and the value of $Q$ (for instance, a lower spot temperature would imply a stronger contrast and thus a smaller area), but the relative variations in the area are largely independent of $c_{\rm s}$ and $Q$ \citep[cf. ][]{Lanzaetal09a}.

%%%%%%%%%%%%%%%%%%%%%%%%%%%%%%%%%%%%%%%%%%%%%%%%%%%%%%%%%%%%%%%%%
%%%%%%%%%%%%%%%%%%%%%%%%%%%%%%%%%%%%%%%%%%%%%%%%%%%%%%

\section{Discussion}
\label{sec_discus}

We find that the rotation of LHS~6343~A is not perfectly synchronized with the orbital motion of the BD companion. Spot modelling is superior to the Lomb-Scargle periodogram for determining the rotation period because a much smaller phase difference is required to distinguish between close periods for a given time baseline. The regular and steady migration of the active longitudes in Fig.~\ref{spot_cont} makes our result quite robust. Unfortunately, the fast  evolution of the individual spots together with the low signal-to-noise ratio of the photometry does not allow us to detect any possible differential rotation. 

The intriguing point concerning the quasi-synchronous rotation of LHS~6343~A is that the tidal interaction between the star and its BD companion is presently very weak. Using the formalism of \citet{Leconteetal10}, we estimate a current timescale for the  tidal angular momentum exchange between the spin and the orbit of $\approx 7.5$~Gyr, if a modified tidal quality factor $Q^{\prime} = 10^{6}$ is adopted for the A component. Assuming that the stellar radius was larger by a factor of $3-4$ during the pre-main-sequence  evolution of the system, the synchronization timescale was shorter by a factor of $\approx 10^{3}$ at that time, i.e., significantly shorter than the pre-main-sequence lifetime of the A component. Therefore, the system should have reached the ZAMS close to synchronization and with the stellar spin aligned with the orbital angular momentum. Then it  evolved to the present state with a likely modest angular momentum loss from the~A component. Indeed,  a reduced angular momentum loss rate in comparison with earlier main-sequence stars seems to be a characteristic of mid-M dwarfs \citep[cf., e.g., ][]{Reinersetal12}. Alternatively, the presence of a close-in companion may have reduced the efficiency of the magnetic braking by the stellar wind, as conjectured by \citet{Lanza10} in the case of stars accompanied by hot Jupiters \citep[cf.  also ][]{Cohenetal10}.

The possibility that magnetic interactions in star-planet systems lead to an enhancement of stellar activity was initially  suggested by \cite{Cuntzetal00} and \cite{Cuntzetal02}.  
Some evidence has been found in several star-planet systems from the modulation of chromospheric activity indexes in phase with the orbit of the planet  rather than with the rotation period of the star \citep{Shkolnik05}. For HD~179949, HD~189733, and $\tau$~Boo, an enhancement of the flux variability has been observed preceding the subplanetary point by $\sim70^{\circ}$ during several observing campaigns \citep{Shkolnik08}.  

In the case of LHS~6343~A, where the companion is a low-mass BD object, the available data are photometric, so the activity modulation is observed in the photosphere. Here, a phase lead of $98^{\circ}$ between the transits and the photometric minima is observed (see Sects.~\ref{sec_obs} and ~\ref{sub_activ}). The effect peaks once per orbit, thus suggesting a magnetic interaction because two relative maxima per orbit are expected in the case of a tidal interaction due to the symmetry of the tidal bulge. Moreover, the relative amplitude of the flux modulation due to the ellipsoidal deformation of LHS~6343~A does not exceed $\sim 5 \times 10^{-5}$ owing to the large separation of the A and C components \citep{Pfahletal08}. Even considering an orbit with an eccentricity of 0.1, the difference in the ellipsoidal light variation between the periastron and the apoastron does not exceed $\sim 3 \times 10^{-5}$ in relative flux units. Therefore, the  observed flux modulation with a relative amplitude of $\sim 10^{-3}$, phased to the orbital motion of the BD, cannot be attributed to the tidal distortion of the primary star.

Similar cases of photospheric active regions rotating in phase with a close-in massive planet have been reported for $\tau$~Boo \citep{Walkeretal08} and CoRoT-4 \citep{Lanzaetal09b}. In the former case, a photospheric active region producing a signature of the order of one millimag was observed in different seasons with the MOST satellite. The feature was  sometimes dark, sometimes bright and was associated in phase with the persistent chromospheric hot spot discussed above. In the case of CoRoT-4, a preferential longitude for spot appearance was found within $\approx 40^{\circ}-50^{\circ}$ from the subcompanion point. The rotation of both stars is close to synchronization as in the case of LHS~6343~A. Although $\tau$~Boo and CoRoT-4 are F-type stars, the same mechanism could be at work in all three cases.  

Several models have been proposed to account for the features of the magnetic interaction between stars and close-in planets. They can be extended to the case of a close BD companion as in LHS~6343~AC. The models by \citet{McIvoretal06}, \citet{Preusse06}, and \citet{Lanza08} have focussed on interpreting of the phase lag between the planet and the maximum visibility of the chromospheric hot spot by adopting different hypotheses, e.g., a tilted potential stellar magnetic field, a finite propagation time of the magnetic perturbation along the  field lines from the planet to the star, or a non-potential twisted field in the stellar corona. They are generally capable of accounting for the phase lag observed in the case of a chromospheric hot spot, but cannot account for an enhancement of the photospheric activity. As shown by the numerical MHD simulations of \citep[][2011a,b]{Cohenetal09}, when the planetary magnetosphere is inside or in touch with the stellar Alfven surface, a remarkable interaction between the stellar and the planetary fields occurs, and the energy released in the magnetic reconnection can be transferred down to the stellar chromosphere and photosphere. This can account for a localized enhancement of the flux, i.e., a local brightening as in large solar flares that have a white-light photospheric counterpart to their chromospheric and coronal emission. However, the  appearance of dark spots associated with a close-by companion requires a different mechanism.

\citet{Lanza08,Lanza2011b} have conjectured that the perturbation of the coronal field and the triggering of flaring activity in the corona may increase the loss rate of magnetic helicity through coronal mass ejections \citep[cf. also ][]{Lanza2012} leading to a perturbation of the stellar dynamo. As reviewed by, e.g., \citet{Brandenburgetal12}, the operation of the dynamo in  late-type stars implies  a mechanism that can get rid of the magnetic helicity that is continuously pumped into the small field scales, otherwise the dynamo is rapidly quenched by magnetohydrodynamic  effects. In the Sun, coronal mass ejections associated with large flares are regarded as a viable mechanism for accomplishing this goal, therefore an enhancement of flaring activity in stars with close-in companions may help the operation of the dynamo. Since the perturbation moves in phase with the planet, the associated loss of magnetic helicity is non-axisymmetric and may lead to a modulation of the $\alpha$ effect of the stellar dynamo in phase with the motion of the companion \citep[see ][ for details]{Lanza08,Lanza2011b}. In systems that are close to synchronization, the modulation can be particularly effective thanks to the slow relative motion of the companion in the reference frame rotating with the star that increases the duration of the interaction between a stellar coronal structure and the magnetic field of the orbiting body. 
The phase lag between the starspots and the companion may be explained by adopting a twisted magnetic field interconnecting the spots with the orbiting BD  as discussed in \citet{Lanza08}. 

An alternative explanation taking into account the destabilizing effects of the tidal perturbation on the magnetic flux tubes inside the red dwarf star cannot of course be excluded. In this scenario, the persistent active longitude would correspond to the periastron passage of the BD companion with the phase lag of $\sim 98^{\circ}$ being the angular distance of that point from the mid-transit. Recent works suggesting a possible gravitational perturbation of the planets on the solar dynamo may give support to this conjecture \citep[cf., e.g., ][ and references therein ]{Abreuetal12}.

\section{Conclusions}
\label{sec_conc}

Considering 508 days of Kepler photometry of the visual system LHS~6343, we have computed a new best fit to the mean photometric transit profile of the C brown dwarf component across the disc
of the A component. We essentially confirm the transit parameters found by  \citet{2011ApJ...730...79J} and  reduce their uncertainties.

A maximum entropy spot model was applied to the Kepler light curve of LHS 6343 to derive maps of the longitudinal distribution of the active regions on the A component, assumed to be responsible for the out-of-transit modulation of the flux.  While including faculae is not found to be critical for obtaining an acceptable fit to the light modulation,   an optimal fit of the light curve is found for a facular-to-spotted area ratio around $Q=8.0$.

Our maps show several active regions with a slower rotation rate ($P_{\rm rot}=13.14 \pm 0.02$~days) than the BD orbital period ($P_{\rm orb}=12.71$~days), implying that the AC system is quasi-synchronized. Owing to the limited signal-to-noise ratio of the data and the fast evolution of the active regions that have lifetimes of $15 20$~days, it is not possible to  detect any differential rotation signature. On the other hand, the starspot covering factor is observed to be systematically enhanced when the spots cross  a longitude leading the substellar point by $\sim100^\circ$. This effect is also detected in the photometric time series as a modulation with the orbital period of the BD showing minima $\sim 3.5$ days before each transit. That this active longitude persists for more than $ \sim 500$ days provides  evidence that the BD companion does affect the photospheric activity of its host somewhat. We conjecture that a  magnetic interaction between the coronal field of the M dwarf and the BD may produce a modulation of the  stellar dynamo leading to an increase in activity modulated in phase with the orbital motion (see Sect.~\ref{sec_discus}). 

A monitoring of the system through measurements of the modulation of the chromospheric lines (e.g., H$\alpha$ or Ca II H\&K) and coronal X-ray emission would be very helpful, because a detection of a hot spot in the upper stellar atmosphere with the same phase lag as the detected photospheric active longitude would indicate interactions through magnetic field reconnection, thus supporting the conjectured mechanism.

\begin{acknowledgements} 
The authors are grateful to an anonymous referee for a careful reading of the manuscript and several valuable suggestions that helped to improve the present paper. 
This work was supported by the /MICINN/ (Spanish Ministry of 
Science and Innovation) - FEDER through grants AYA2009-06934, AYA2009-14648-C02-01, 
and CONSOLIDER CSD2007-00050. E.~Herrero is supported by a JAE Pre-Doc grant (CSIC). 
\end{acknowledgements}

\bibliographystyle{aa} % style aa.bst

\appendix

\section{Influence of the adopted model parameters on the spot maps}
\label{appendixa}

\subsection{Time resolution ($\Delta t_{\rm f}$) and amount of regularization ($\beta$)}

In Sect.~\ref{sec_params} we discussed the method applied to determining the maximum time interval $\Delta t_{\rm f}$ that our model can accurately fit when considering a fixed distribution of three active regions \citep{Lanzaetal03}. This corresponds to the time resolution of our model, since we assume that active regions do not evolve on timescales shorter than $\Delta t_{\rm f}$. The total interval $T$ was divided to $N_{\rm f}$ segments so that $\Delta t_{\rm f}=T/N_{\rm f}$. The optimal $N_{\rm f}$ was found to be 58 from the best fit of the three-spot model, as measured from the $\chi^{2}$ statistics. 

\begin{figure}[]
\centerline{
\includegraphics[height=9.5cm,angle=-90]{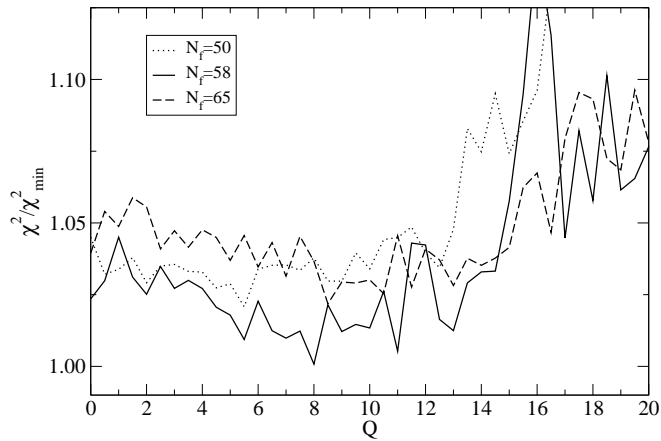}}
\caption{The ratio of the $\chi^{2}$ of the three-spot model fit to the minimum value for $N_{f}=58$ ($\chi^{2}_{\rm min}$) vs. the parameter $Q$, i.e., the ratio of the facular area to the cool spot area in active regions. Solution series for three different values of the time interval $\Delta t_{\rm f}=T/N_{f}$ are shown.
}
\label{qchi_all}
\end{figure}

\begin{figure}[]
\centerline{
\includegraphics[height=9.5cm,angle=90]{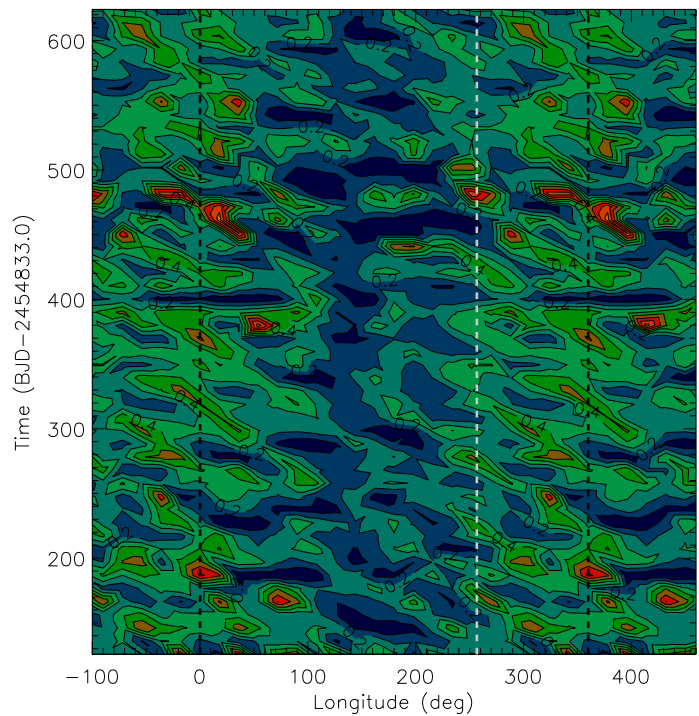}}  
\caption{As in Fig.~\ref{spot_cont}, but adopting a longer time interval $\Delta t_{\rm f} = 10.1719$~days.}
%Isocontours of the ratio $f/f_{\rm max}$, where $f$ is the spot covering factor and $f_{\rm max}= 0.00032$ its maximum value,  versus time and longitude for the ME models with $Q=8.0$. The colour coding adopted to indicate different filling factor levels  is the same as in Fig.~\ref{spot_cont}. The dashed white line marks the longitude of the sub-companion point, which is fixed in this reference frame at $\sim257^{\circ}$. A longer time interval of $\Delta t_{\rm f}=10.1719$ was used in the regularized ME model.}
\label{sc_n50}
\end{figure}

Three cases corresponding to different lengths of the time interval $\Delta t_{\rm f}$ are presented in Fig.~\ref{qchi_all}, where the ratio $\chi^{2}/ \chi^{2}_{\rm min}$ is plotted vs. $Q$, the facular-to-spotted area ratio; $\chi^{2}$ is the chi square of the composite best fit to the total time series; and $\chi^{2}_{\rm min}$ is the minimum $\chi^{2}$ obtained for $N_{f}=58$ and $Q=8.0$.  The corresponding spot model is discussed in the text.
Here we  explore the case with $N_{\rm f}=50$ and perform an ME analysis with the corresponding time interval, i.e., $\Delta t_{\rm f}=10.1719$ days, to investigate the effect of a different $\Delta t_{\rm f}$ on our results.
The same approach as described in Sect.~\ref{sub_reslc} was considered, and regularized ME models were iteratively adjusted until $\beta=2$. The isocontour map of the distribution of the spot filling factor vs. time and longitude is presented in Fig.~\ref{sc_n50}. It is remarkably similar to the one presented in Fig.~\ref{spot_cont} and the phased activity enhancement around $\sim 350^{\circ}$ is still visible, although the migration of the starspots and their changes on short timescales are less evident. Therefore, a longer time interval limits the possibility of accurately measuring spots' migration and estimate their lifetimes.

\begin{figure}[]
\centerline{
\includegraphics[height=9.5cm,angle=90]{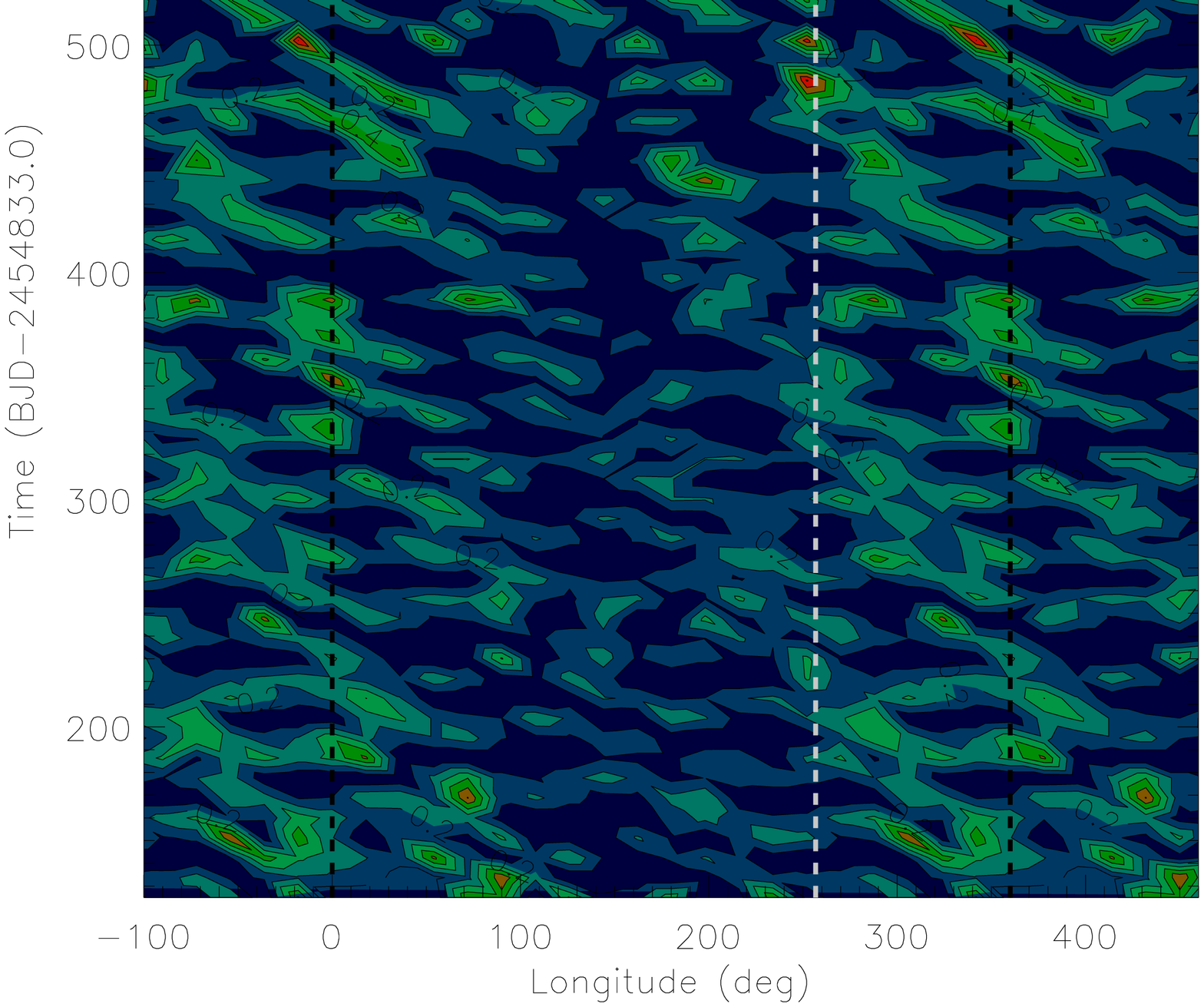}} 
\caption{As in Fig.~\ref{spot_cont}, but performing an ME regularization with $\beta=1$.}
%Isocontours of the ratio $f/f_{\rm max}$, where $f$ is the spot covering factor and $f_{\rm max}= 0.00061$ its maximum value,  versus time and longitude for the ME models with $Q=8.0$.  The colour coding adopted to indicate different filling factor levels  is the same as in Fig.~\ref{spot_cont}. The dashed white line marks the longitude of the sub-companion point, which is fixed in this reference frame at $\sim257^{\circ}$. The ME model is regularized with $\beta=1$. }
\label{sc_b1}
\end{figure}

In a second test, we chose the same time interval as in Sect.~\ref{sec_results}, i.e.,  $N_{\rm f}=58$,  to explore the effects of a different amount of regularization on the ME spot maps.   In Fig.~\ref{sc_b1}, we present a spot map obtained with a lower regularization ($\beta=1$). At that point, the large starspots appearing in  Fig.~\ref{spot_cont} are resolved into several smaller spots.  This fine structure changes vs. the time without any clear regularity, as expected for an artefact due to  overfitting  the noise present in the data. 

\begin{figure}[]
\centerline{
\includegraphics[height=9.5cm,angle=90]{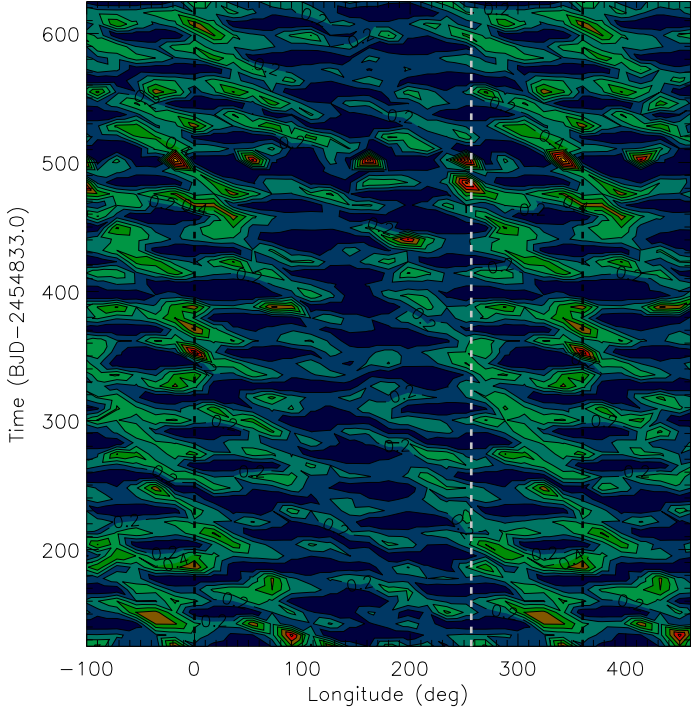}} 
\caption{As in Fig.~\ref{spot_cont}, but performing an ME regularization with $\beta = 3$.}
%Isocontours of the ratio $f/f_{\rm max}$, where $f$ is the spot covering factor and $f_{\rm max}= $ ??? [{\bf Kike, please put the right maximum value here}] its maximum value,  versus time and longitude for the ME models with $Q=8.0$. The colour coding adopted to indicate different filling factor levels  is the same as in Fig.~\ref{spot_cont}. The dashed white line marks the longitude of the sub-companion point, which is fixed in this reference frame at $\sim257^{\circ}$. The ME model is regularized with $\beta=3$. [{\bf The dashed vertical black lines at 0 and 360 degrees are missing. }]}
\label{sc_b3}
\end{figure}

Finally, we plot the distribution of the filling factor obtained with $\beta =3$ in Fig.~\ref{sc_b3}. The effects of the overegularization described in Sect.~\ref{sub_reslc} can be seen as a tendency for the appearance of larger and smoother spot groups that include previously resolved starspots. Three main groups separated by approximately $120^{\circ}$ in longitude generally appear, showing the same backwards migration as with $\beta=2$, the optimal value adopted for the regularization. The residuals of the composite best fit to the light curve (not shown) show larger systematic deviations than in the cases with $\beta=1$ or $\beta=2$, again a sign of overegularization. 

%%%%%%%%%%%%%%%%%%%%%%%%%%%%%%%%%%%%%%%%%%%%%%%%%%%%%%%%%%%%%%%%%
%%%%%%%%%%%%%%%%%%%%%%%%%%%%%%%%%%%%%%%%%%%%%%%%%%%%%%%%%%%%%%%%%

%%%%%%%%%%%%%%%%%%%%%%%%%%%%%%%%%%%%%%%%%%%%%%%%%%%%%%%%%%%%%%%%%
%%%%%%%%%%%%%%%%%%%%%%%%%%%%%%%%%%%%%%%%%%%%%%%%%%%%%%%%%%%%%%%%%

%%%%%%%%%%%%%%%%%%%%%%%%%%%%%%%%%%%%%%%%%%%%%%%%%%%%%%%%%%%%%%%%%
%%%%%%%%%%%%%%%%%%%%%%%%%%%%%%%%%%%%%%%%%%%%%%%%%%%%%%%%%%%%%%%%%

%%%%%%%%%%%%%%%%%%%%%%%%%%%%%%%%%%%%%%%%%%%%%%%%%%%%%%%%%%%%%%%%%
%%%%%%%%%%%%%%%%%%%%%%%%%%%%%%%%%%%%%%%%%%%%%%%%%%%%%%%%%%%%%%%%%

\subsection{Inclination of the stellar spin axis}

The inclination of the stellar spin  is assumed to be equal to that of the orbit of the BD companion in our analysis, i.e. $i=89 \fdg 60$. An isotropic orientation of the stellar spin axis has its mode at $i=90^{\circ}$ \citep{Herrero12}, so it is reasonable to choose a value close to this, given that there is no a priori information to constrain the parameter.

The tidal timescale for the alignment of the stellar spin and the orbital angular momentum are comparable, in the case of LHS~6343~A, to the synchronization timescale that is estimated to be  $\approx 7.5$~Gyr. Given a probable age of the system between 1 and 5 Gyr, we cannot be sure that the obliquity has been damped by tides, unless we assume that most of the damping occurred during the pre-main-sequence phase when the tidal interaction was stronger due to a larger $R_{\rm A}$ (cf. Sect.~\ref{sec_discus}). Therefore, we adopt a prudential approach and explore  different cases in which the system is out of spin-orbit alignment.

\begin{figure}[]
\centerline{
\includegraphics[height=9.5cm,angle=90]{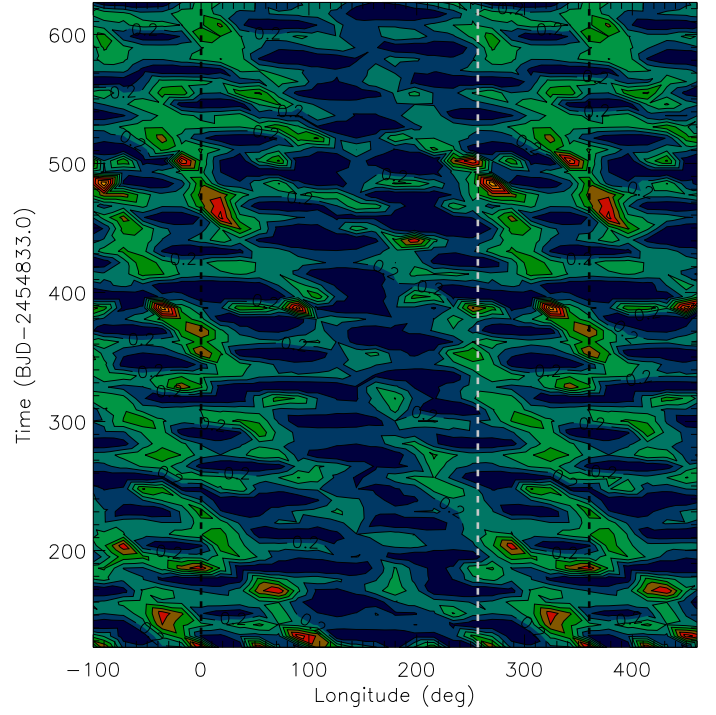}} 
\caption{As in Fig.~\ref{spot_cont}, but adopting a stellar inclination of $i=60^{\circ}$.}
%Isocontours of the ratio $f/f_{\rm max}$, where $f$ is the spot covering factor and $f_{\rm max}= 0.00035$ its maximum value,  versus time and longitude for the ME models with $Q=8.0$.  The colour coding adopted to indicate different filling factor levels  is the same as in Fig.~\ref{spot_cont}. The dashed white line marks the longitude of the sub-companion point, which is fixed in this reference frame at $\sim257^{\circ}$. A stellar inclination of $i=60^{\circ}$ was adopted in this case.}
\label{sc_i60}
\end{figure}

\begin{figure}[]
\centerline{
\includegraphics[height=9.5cm,angle=90]{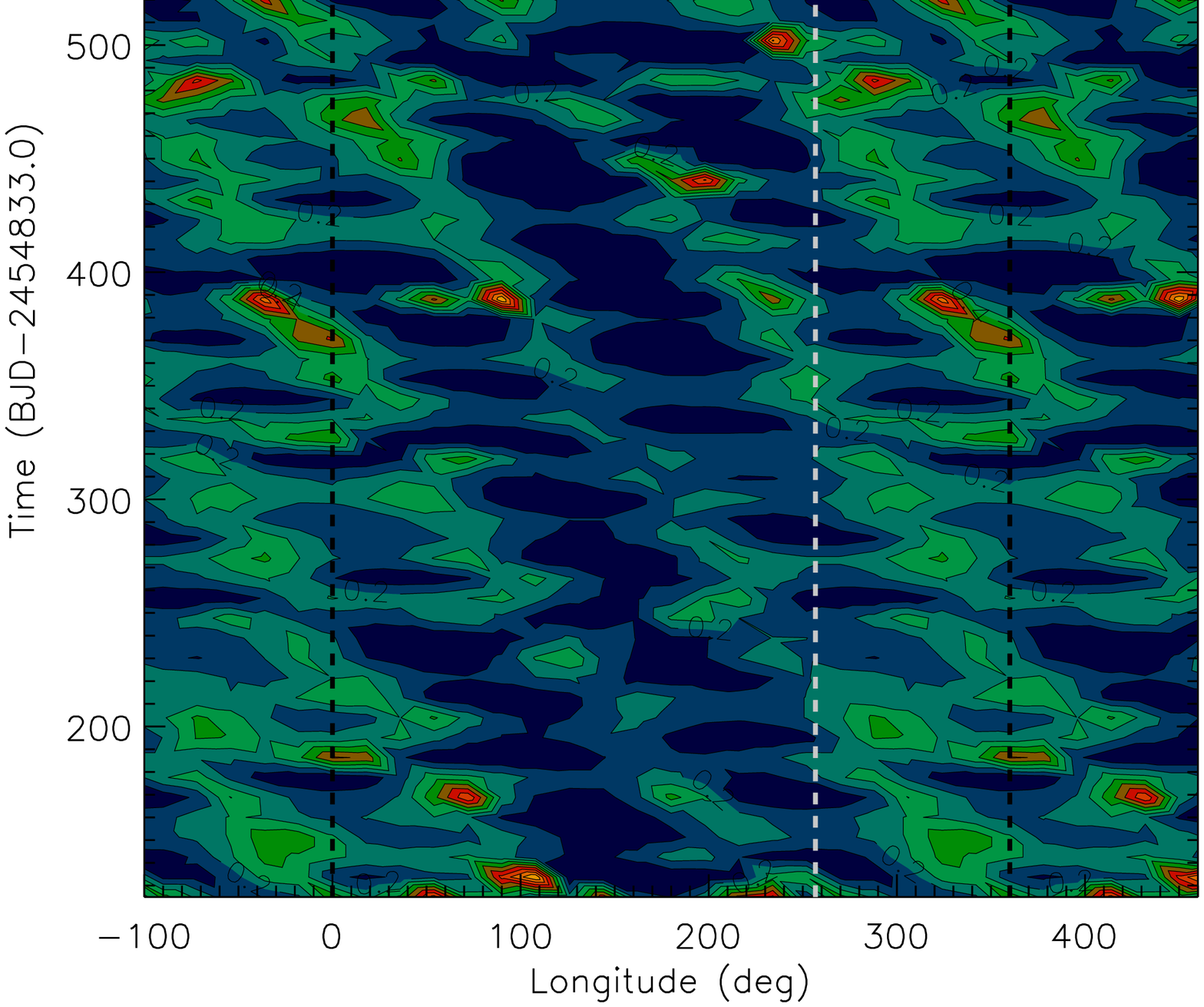}} 
\caption{As in Fig.~\ref{spot_cont}, but adopting a stellar inclination of $i=45^{\circ}$.}
%Isocontours of the ratio $f/f_{\rm max}$, where $f$ is the spot covering factor and $f_{\rm max}= 0.00033$ its maximum value,  versus time and longitude for the ME models with $Q=8.0$. The colour coding adopted to indicate different filling factor levels  is the same as in Fig.~\ref{spot_cont}. The dashed white line marks the longitude of the sub-companion point, which is fixed in this reference frame at $\sim257^{\circ}$. A stellar inclination of $i=45^{\circ}$ was adopted in this case. }
\label{sc_i45}
\end{figure}

We adopt the same parameters as in Sect.~\ref{sub_reslc} and compute a regularized ME best fit of the light curve with $\beta=2$, fixing the stellar inclination to $i=60^{\circ}$ and $i=45^{\circ}$. The resulting spot maps are plotted in Figs.~\ref{sc_i60} and \ref{sc_i45}, respectively. We may expect that the latitudinal distribution of the spots are remarkably different, especially for $i=45^{\circ}$. On the other hand, the plotted distributions of the filling factor vs. longitude and time are  similar to what is computed with $i=89\fdg 60$. Also the evolution and migration of the  active regions are similar to that case. Only some details appear to be critically dependent on the adopted inclination. 

The results of the tests described in this Appendix confirms that by averaging the ME spot maps over latitude, we can effectively remove most of the degeneracy present in the light curve inversion process.

%%%%%%%%%%%%%%%%%%%%%%%%%%%%%%%%%%%%%%%%%%%%%%%%%%%%%%%%%%%%%%%%%
%%%%%%%%%%%%%%%%%%%%%%%%%%%%%%%%%%%%%%%%%%%%%%%%%%%%%%%%%%%%%%%%%

%%%%%%%%%%%%%%%%%%%%%%%%%%%%%%%%%%%%%%%%%%%%%%%%%%%%%%%%%%%%%%%%%
%%%%%%%%%%%%%%%%%%%%%%%%%%%%%%%%%%%%%%%%%%%%%%%%%%%%%%%%%%%%%%%%%

\end{document}